\documentclass[runningheads]{llncs}
\usepackage[margin=1in]{geometry} 
\usepackage{graphicx} 
\usepackage{booktabs}
\usepackage{xcolor}
\newcommand{\rb}[1]{\textbf{\textcolor{red}{#1}}}
\usepackage{multirow}
\usepackage{subcaption} 
\usepackage{enumitem}
\usepackage[most]{tcolorbox}
\usepackage{hyperref}
\usepackage{pifont}
\usepackage{amsmath,amssymb,amsfonts}
\usepackage{algorithm}
\usepackage[noend]{algpseudocode}
\usepackage{float}

\newcommand{\cmark}{\ding{51}}

\newcommand{\OURS}{OASIS}
\title{OASIS: Optimizing Attacker Sequences for Hard-Label Black-Box Text Attacks}
\date{}

\author{
  Qian Chen$^{*}$
  \and
  Shiliang Xiao$^{*}$
  \and
  Yuzhi Liang$^{\dagger}$
}

\institute{
  School of Information Science and Technology\\
  Guangdong University of Foreign Studies, China\\
  \email{20241050023@mail.gdufs.edu.cn, slxiao@mail.gdufs.edu.cn, yzliang@gdufs.edu.cn}\\[0.5ex]
  $^{*}$Equal contribution.
  \quad
  $^{\dagger}$Corresponding author.
}

\titlerunning{OASIS}
\authorrunning{Q. Chen et al.}

\begin{document}

\maketitle

\begin{abstract}
Different attack methods follow different search trajectories, they succeed on different subsets of samples, whereas existing hard-label black-box text attacks mainly focus on improving individual attackers or manually combining them. We present {\OURS}, a method for optimizing attacker sequences in hard-label black-box text attacks. {\OURS} first performs a one-time bi-objective attack chain search over candidate sequences to balance attack success rate and perturbation, and then reuses the selected fixed global chain during attack chain execution. Experiments across multiple datasets, victim models, and large language models show that {\OURS} consistently outperforms strong standalone baselines and simple manually constructed chains. These results suggest that attacker composition is not merely an implementation choice, but a practical optimization target for improving hard-label black-box text attacks.
\end{abstract}

\section{Introduction}
Deep neural networks have achieved strong performance across a wide range of NLP tasks, yet they remain vulnerable to adversarial examples. In realistic settings, target models often expose no intermediate information and operate under limited query budgets. Hard-label black-box text attacks aim to induce misclassification by making only a small number of perturbations to the original text while preserving its semantics as much as possible, given only label-only feedback. Different attack methods follow different search trajectories, and therefore their successful attack sets do not fully overlap. This raises a central challenge: \textit{how can we search for an attack chain that exploits cross-attacker complementarity so as to maximize the overall coverage of successful attacks?}

Existing hard-label black-box text attacks mainly focus on designing stronger standalone attackers or manually combining attack methods. Methods such as LeapAttack~\cite{DBLP:conf/kdd/YeCMWM22} and TextHoaxer~\cite{ye2022texthoaxer} seek better attack directions under label-only feedback, while LimeAttack~\cite{DBLP:conf/aaai/ZhuZ0WL24}, VIWHard~\cite{DBLP:journals/ijon/ZhangWGZWL25}, and TextHacker~\cite{DBLP:conf/emnlp/YuWC022} improve token ranking, surrogate modeling, or local search.  Meanwhile, empirical hand-crafted attack combinations such as OpenFact~\cite{DBLP:journals/corr/abs-2409-02649} do not treat attacker complementarity as an explicit optimization target. Because different attackers follow different search trajectories, they succeed on different subsets of samples. Moreover, strong attackers often exhibit larger overlap in their successful attack sets, whereas relatively weaker attackers may cover residual cases missed by stronger ones. \textbf{Overall}, most existing methods either focus on improving individual attackers or rely on manually combining them, without explicitly exploiting the complementarity among attackers induced by different search trajectories.

To address this issue, we propose {\OURS}\footnote{Our code is anonymously available at \url{https://github.com/s1xiao/OASIS}.}, a method for \textbf{O}ptimizing \textbf{A}ttacker \textbf{S}equences \textbf{i}n hard-label black-box text attack\textbf{S}. Specifically, our method consists of two stages: local chain search and attack chain execution. On a local model and dataset, we perform a one-time search over candidate attack chains, jointly maximizing ASR and minimizing Pert as a bi-objective optimization problem. We use NSGA-II to search for the Pareto front, and then define the positive ideal solution and negative ideal solution in this solution space to select a fixed global attack chain via TOPSIS. On target models and datasets, the same global attack chain is reused and executed sequentially on each sample.

Extensive experiments across multiple target datasets, target models, and large language models show that {\OURS} consistently outperforms strong standalone baselines and manual compositions, while the selected attack chain remains effective beyond the local search setting.

Our main contributions are as follows:
\begin{itemize}[leftmargin=*, noitemsep, topsep=2pt]
    \item We identify that different attackers follow different search trajectories and therefore cover only partially overlapping subsets of successful samples.
    \item We propose a global attack-chain search framework that selects a fixed reusable attack chain by capturing cross-attacker complementarity.
    \item Experiments across target datasets and victim models show that the searched chain consistently outperforms strong standalone attackers and manually constructed chains.
\end{itemize}

\section{Related Work}

Research on textual adversarial attacks is formally categorized based on the adversary's knowledge of the victim model.

\paragraph{White-box Attacks.}
When richer feedback is available, adversarial text generation can rely on gradients or confidence scores for more direct optimization. In the white-box setting, HotFlip~\cite{DBLP:conf/acl/EbrahimiRLD18} performs gradient-based character perturbations, GBDA~\cite{DBLP:conf/emnlp/GuoSJK21} optimizes adversarial distributions through Gumbel-Softmax relaxation, and TextGrad~\cite{DBLP:conf/iclr/HouJZZ00C23} further exploits gradient signals for robustness-oriented text optimization. Some early attacks also combine perturbations across different granularities: TextBugger~\cite{DBLP:conf/ndss/LiJDLW19} integrates character- and word-level modifications, while DeepWordBug~\cite{DBLP:conf/sp/GaoLSQ18} uses scoring-based character manipulations. In soft-label black-box settings, where gradients are unavailable but confidence scores can still guide search, representative methods mainly differ in how they select and generate substitutions. Many substitution-based attacks are built on lexical resources such as WordNet~\cite{DBLP:journals/cacm/Miller95} and HowNet~\cite{dong2006hownet}. PWWS~\cite{ren-etal-2019-generating}, PSOAttack~\cite{DBLP:conf/acl/ZangQYLZLS20}, and TextFooler~\cite{JinJZS20} focus on saliency-guided or search-based word substitution, whereas BERT-Attack~\cite{DBLP:conf/emnlp/LiMGXQ20}, BAE~\cite{DBLP:conf/emnlp/GargR20}, and CLARE~\cite{DBLP:conf/naacl/LiZPCBSD21} improve substitution quality with masked language models and richer edit operations. Beyond direct substitution, ATGSL~\cite{DBLP:conf/emnlp/LiSLKWZHL23} leverages conditional generation to produce fluent adversarial examples, and ALGEN~\cite{DBLP:conf/acl/0002XB25} studies few-shot embedding inversion via cross-model alignment. The development of this line of work has also been supported by toolkit efforts such as TextAttack~\cite{DBLP:conf/emnlp/MorrisLYGJQ20}, which systematizes adversarial attack recipes in NLP.

\paragraph{Black-box Attacks.}
Hard-label black-box attacks are more restrictive because the attacker observes only the final predicted label. Existing work in this setting mainly improves the search procedure of a \textbf{single attacker}, although the technical routes differ. Population-based methods such as HLBB~\cite{DBLP:conf/aaai/MaheshwaryMP21} adopt genetic search for hard-label attack generation, while HyGloadAttack~\cite{DBLP:journals/nn/LiuXLYLZX24} introduces hybrid optimization to alleviate local optima during label-only search. Boundary-approximation and direction-guided methods instead focus on how to approach the decision boundary more efficiently: GeoAttack~\cite{DBLP:conf/coling/MengW20} exploits geometric structure to approximate effective perturbation directions, LeapAttack~\cite{DBLP:conf/kdd/YeCMWM22} further explores the decision boundary through directional cues under label-only feedback, and TextHoaxer~\cite{ye2022texthoaxer} refines search directions in continuous embedding space for hard-label attacks. Another line of work improves local token selection and perturbation refinement under discrete constraints: SemAttack~\cite{DBLP:conf/naacl/WangXLCL22} emphasizes semantic consistency during adversarial generation, TextHacker~\cite{DBLP:conf/emnlp/YuWC022} improves local search with token-level perturbation heuristics and attack-history feedback, LimeAttack~\cite{DBLP:conf/aaai/ZhuZ0WL24} estimates token importance with local surrogate signals and combines them with beam search, and VIWHard~\cite{DBLP:journals/ijon/ZhangWGZWL25} identifies important words with masked language models to guide natural substitutions. This direction is also related to model-agnostic explanation methods such as Anchors~\cite{DBLP:conf/aaai/Ribeiro0G18} and to constrained synonym spaces built from counter-fitted word vectors~\cite{mrksic2016counterfitting}. More recent methods also place greater emphasis on attack quality under hard-label feedback: HQAAttack~\cite{DBLP:journals/corr/abs-2402-01806} focuses on query-efficient generation of high-quality adversarial texts, and SSPAttack~\cite{DBLP:conf/aaai/0008X0XZMC0Z23} strengthens substitution-based attacks with structure-aware perturbation design. Beyond single-attacker optimization, some empirical attack pipelines also combine multiple methods; OpenFact~\cite{DBLP:journals/corr/abs-2409-02649} manually integrates attack algorithms for fact-checking tasks. However, such combinations do not explicitly optimize attacker complementarity or the order of attacker execution. 

\section{Methodology}
\begin{figure}[t]
  \centering
  \includegraphics[width=1\linewidth]{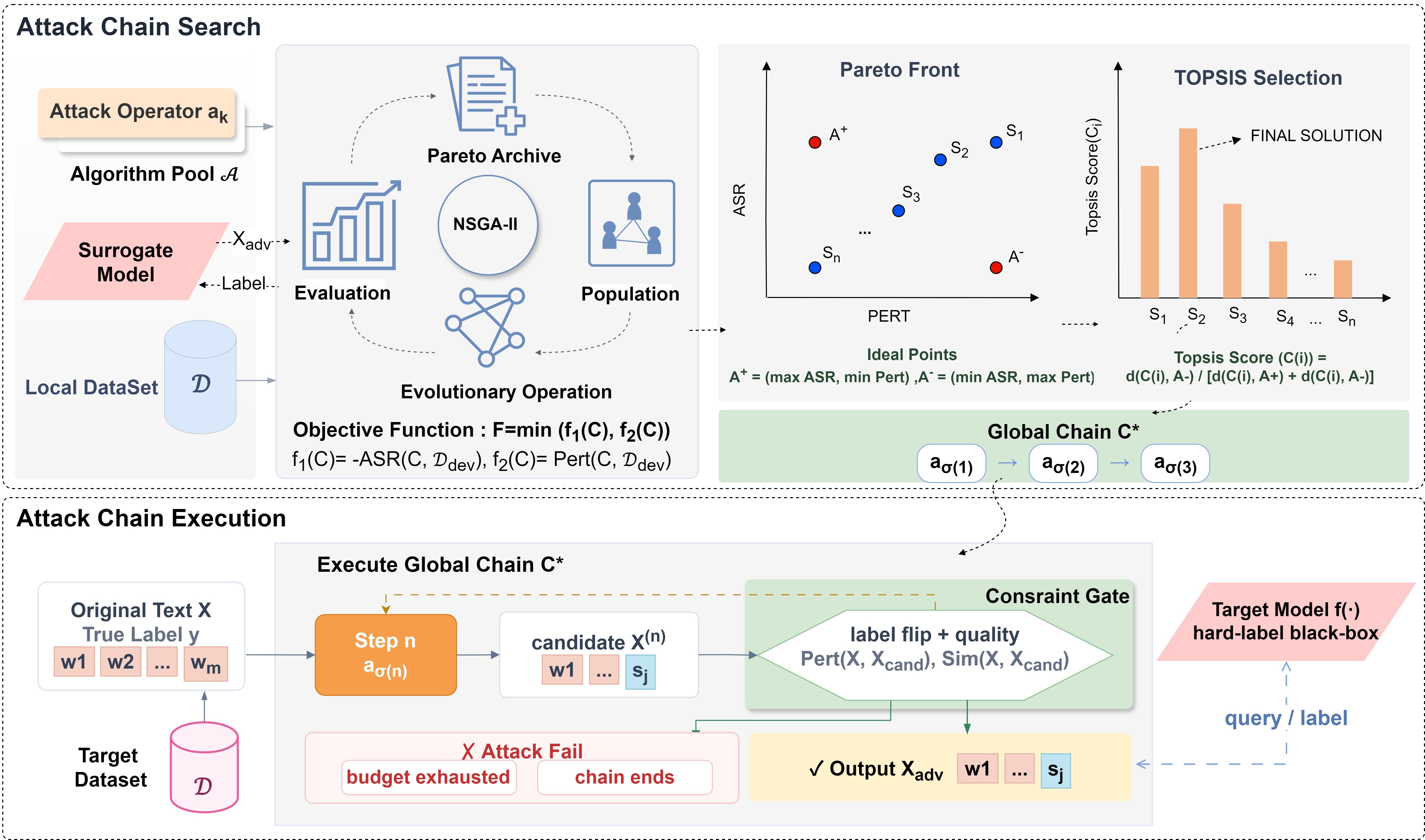}
  \caption{Overall framework of {\OURS}. The method first performs attack chain search on the local dataset under the local model to identify an effective fixed global chain using NSGA-II and TOPSIS, and then reuses the selected chain on target datasets and target models under query budget and quality constraints.}
  \label{fig:pic_overview}
\end{figure}

Before introducing the algorithm in detail, we first clarify an important intuition: {\OURS} does not aim to directly select the attacker combination with the highest standalone performance in the search space. Instead, since different attackers succeed on different subsets of samples, {\OURS} explicitly exploits cross-attacker complementarity and identifies a chain whose components work well together, thereby improving the overall coverage of successful attacks. The overall workflow is illustrated in Figure~\ref{fig:pic_overview}.

Specifically, {\OURS} consists of two stages: (1) \textbf{Attack Chain Search}, where candidate chains are evaluated on a local dataset under a local model and optimized under a joint ASR--Pert objective; and (2) \textbf{Attack Chain Execution}, where the selected fixed global chain is reused across target datasets and target models under a limited query budget and unified quality constraints.

\subsection{Attack Chain Formulation}

\noindent\textbf{Problem definition.}
Given a text classifier $f: \mathcal{X} \rightarrow \mathcal{Y}$, an input $X = [w_1, w_2, \ldots, w_n]$ with label $y$, a query budget $B$, a perturbation threshold $\rho$, and a semantic similarity threshold $\tau$, the goal in the hard-label black-box setting is to find an adversarial example $X'$ such that
\begin{equation}
  f(X') \neq y,\quad
  \mathrm{Pert}(X, X') \leq \rho,\quad
  \mathrm{Sim}(X, X') \geq \tau,\quad
  Q_{\mathrm{used}} \leq B
  \label{eq:adv_obj}
\end{equation}
where $\mathrm{Pert}(X, X') = \frac{1}{n}\sum_{i=1}^{n}\mathbf{1}(w_i \neq w_i')$, $\mathrm{Sim}(X, X')$ is measured by the same sentence-level similarity module used in the quality check during attack execution, and $Q_{\mathrm{used}}$ is the total number of victim-model queries. We evaluate attacks only on originally correct samples satisfying $f(X)=y$.

\noindent\textbf{Attack chain.}
Standalone performance does not determine chain performance, so we optimize the ordered attacker sequence directly. Let the attacker pool be $\mathcal{A} = \{a_1, a_2, \ldots, a_K\}$, where $K=7$. The attack chain is defined as an ordered sequence without repetition:
\begin{equation}
  C = \bigl(a_{\sigma(1)},\, a_{\sigma(2)},\, \ldots,\, a_{\sigma(L)}\bigr),
  \quad \sigma:\{1,\ldots,L\}\to\{1,\ldots,K\}\ \text{injective}.
  \label{eq:chain}
\end{equation}
The candidate space is denoted by $\mathcal{C}_L$, with cardinality $|\mathcal{C}_L| = P(K,L) = K!/(K-L)!$. During execution, attackers are invoked sequentially on the same original input $X$. At step $i$, the $i$-th attacker receives the original text, the gold label, and the remaining budget, and produces
\begin{equation}
  \bigl(X_{\mathrm{cand}}^{(i)},\, s_i,\, q_i\bigr)
  = a_{\sigma(i)}\bigl(X, y, B_{\mathrm{rem}}^{(i)}\bigr),\quad
  B_{\mathrm{rem}}^{(i)} = B - \sum_{j<i} q_j,
  \label{eq:chain_step}
\end{equation}
where $X_{\mathrm{cand}}^{(i)}$ is the returned candidate text, $s_i \in \{0,1\}$ indicates whether the attacker finds a label-flipping candidate, and $q_i$ is the number of queries consumed at that step. If the candidate is empty or violates the quality constraints, the chain continues to the next attacker. All candidates are checked against the same original input $X$.

To evaluate one chain on the local development set, let $\mathcal{D}_{L}^{\mathrm{corr}} = \{(X,y)\in\mathcal{D}_{L}: f(X)=y\}$ be the subset of originally correct samples. Let $s(X;C)\in\{0,1\}$ denote whether chain $C$ returns a valid adversarial example for sample $X$, and let $\mathcal{S}(C)=\{X\in\mathcal{D}_{L}^{\mathrm{corr}}: s(X;C)=1\}$ be the successful subset. We then define
\begin{equation}
  \mathrm{ASR}_{L}(C)=
  \frac{1}{|\mathcal{D}_{L}^{\mathrm{corr}}|}
  \sum_{X\in\mathcal{D}_{L}^{\mathrm{corr}}} s(X;C),
  \label{eq:asr_dev}
\end{equation}
\begin{equation}
  \mathrm{Pert}_{L}(C)=
  \frac{1}{|\mathcal{S}(C)|}
  \sum_{X\in\mathcal{S}(C)}
  \mathrm{Pert}\bigl(X, X_{\mathrm{adv}}(X;C)\bigr),
  \label{eq:pert_dev}
\end{equation}
where $X_{\mathrm{adv}}(X;C)$ denotes the accepted adversarial output returned by $C$ on $X$. If $\mathcal{S}(C)=\varnothing$, we assign the worst perturbation score $\mathrm{Pert}_{L}(C)=1$. {\OURS} therefore aims to find a fixed reusable chain that jointly attains high $\mathrm{ASR}_{L}$ and low $\mathrm{Pert}_{L}$.

\subsection{Attack Chain Search via NSGA-II}

Exhaustively evaluating all $P(K,L)$ candidate chains is impractical because each candidate requires real attack queries on the victim model. We therefore perform a one-time search with NSGA-II(\cite{DBLP:journals/tec/DebAPM02}) on the local dataset under the local model, using a small development set $\mathcal{D}_{\mathrm{dev}}$ with $N_{\mathrm{eval}}=100$ samples, and then reuse the selected chain across target datasets and target models.

The search stage solves the following bi-objective problem:
\begin{equation}
  \min_{C\in\mathcal{C}_L}\ \mathbf{f}(C)
  = \bigl(-\mathrm{ASR}_{L}(C),\ \mathrm{Pert}_{L}(C)\bigr).
  \label{eq:search_objective}
\end{equation}

Let $\mathcal{P}_t \subseteq \mathcal{C}_L$ denote the population at generation $t$, where each individual is a valid attack chain of length $L$. We initialize $\mathcal{P}_0$ by sampling valid non-repeated chains from $\mathcal{C}_L$ and evaluate every chain under the same budget and quality constraints as in final testing.

Given two parent chains
\[
C^{(1)}=(a^{(1)}_1,\ldots,a^{(1)}_L), \qquad
C^{(2)}=(a^{(2)}_1,\ldots,a^{(2)}_L),
\]
we generate offspring by an order-preserving crossover: a crossover point $r \in \{1,\ldots,L-1\}$ is first sampled, the offspring inherits the prefix $(a^{(1)}_1,\ldots,a^{(1)}_r)$ from the first parent, and the remaining positions are filled by scanning the second parent from left to right and appending attackers that have not yet appeared in the inherited prefix. In this way, crossover is defined directly on ordered attacker chains and preserves both order information and the no-repetition constraint as much as possible.

We then apply a chain-level mutation to each offspring. With probability $p_m$, we either swap two positions in the chain or replace one position with an attacker that is not currently used in that chain. The swap operator changes the execution order of attackers, whereas the replace operator changes the attacker composition itself. After crossover and mutation, duplicated decoded chains are removed. If an offspring contains repeated attackers, it is treated as invalid.

Let $\mathcal{O}_t$ denote the offspring set generated from $\mathcal{P}_{t-1}$. NSGA-II then selects the next population from the combined parent--offspring set:
\begin{equation}
  \mathcal{P}_t
  =
  \operatorname{Select}_{\textsc{NSGA-II}}
  \bigl(\mathcal{P}_{t-1}\cup \mathcal{O}_t;\, P,\mathbf{f}\bigr).
  \label{eq:nsgaii_select}
\end{equation}
After $G$ generations, we obtain the final Pareto front $\mathcal{F}_{\mathrm{pareto}}$, from which TOPSIS selects a single compromise chain for later attack chain execution. The overall search procedure is shown in Algorithm~\ref{alg:chain-search-main}.

\begin{algorithm}[t]
\caption{Attack-Chain Search}
\label{alg:chain-search-main}
\begin{algorithmic}[1]
\State $\mathcal{P}_0 \gets$ sample $P$ valid chains from $\mathcal{C}_L$
\State Evaluate each $C \in \mathcal{P}_0$ and compute $\mathbf{f}(C)$
\For{$t = 1,2,\ldots,G$}
    \State Select parent pairs from $\mathcal{P}_{t-1}$
    \State Generate offspring $\mathcal{O}_t$ by order-preserving crossover
    \State Apply swap-or-replace mutation to each offspring
    \State Remove duplicated chains and penalize invalid repeated chains
    \State Evaluate each valid $C \in \mathcal{O}_t$ and compute $\mathbf{f}(C)$
    \State $\mathcal{P}_t \gets \operatorname{Select}_{\textsc{NSGA-II}}(\mathcal{P}_{t-1}\cup\mathcal{O}_t;\, P,\mathbf{f})$
\EndFor
\State Extract the final Pareto front $\mathcal{F}_{\mathrm{pareto}}$ from $\mathcal{P}_G$
\end{algorithmic}
\end{algorithm}

\subsection{TOPSIS-based Chain Selection}

After NSGA-II terminates, the Pareto front $\mathcal{F}_{\mathrm{pareto}}$ contains multiple candidate chains with different ASR--Pert trade-offs. We use TOPSIS~\cite{DBLP:journals/eswa/BehzadianOYI12} (Technique for Order Preference by Similarity to Ideal Solution) to select a single compromise solution, which is used as the fixed global chain in {\OURS}.

Let the candidate set be $\mathcal{F}_{\mathrm{pareto}} = \{C^{(1)}, \ldots, C^{(m)}\}$, where each chain is associated with a score pair $(\mathrm{ASR}_{L}(C^{(i)}),\, \mathrm{Pert}_{L}(C^{(i)}))$. We first normalize these attributes to form a decision matrix, then define the positive ideal solution as $A^+ = (\max_i \mathrm{ASR}_{L}(C^{(i)}),\; \min_i \mathrm{Pert}_{L}(C^{(i)}))$ and the negative ideal solution as $A^- = (\min_i \mathrm{ASR}_{L}(C^{(i)}),\; \max_i \mathrm{Pert}_{L}(C^{(i)}))$. Let $d(\cdot,\cdot)$ denote Euclidean distance in the normalized two-objective space. The selected chain is the one with the highest relative closeness:
\begin{equation}
  C^* = \mathop{\arg\max}_{C^{(i)} \in \mathcal{F}_{\mathrm{pareto}}}
        \frac{d(C^{(i)}, A^-)}{d(C^{(i)}, A^+) + d(C^{(i)}, A^-)}
  \label{eq:topsis}
\end{equation}

The purpose of this selection step is not to pursue an extreme optimum on one metric, but to obtain a globally reusable chain with a more balanced ASR--Pert trade-off. In our implementation, this amounts to selecting the Pareto-front solution that is closest to the normalized ideal point of high ASR and low Pert. As a result, TOPSIS naturally disfavors frontier endpoints that look strong only because they over-optimize one objective on the local dataset under the local model, and instead prefers compromise chains that remain competitive on both metrics.

\subsection{Attack Chain Execution}

For each test sample $X$, we first query the victim model. If $f(X) \neq y$, the sample is skipped and is not counted as a valid attack target. Otherwise, the fixed chain $C^* = (a_{\sigma(1)}, \ldots, a_{\sigma(L)})$ is executed sequentially under the shared budget. At step $i$, attacker $a_{\sigma(i)}$ is invoked according to Eq.~\eqref{eq:chain_step}, and the total query count is updated as $Q_{\mathrm{used}} \leftarrow Q_{\mathrm{used}} + q_i$. If $Q_{\mathrm{used}} \ge B$, the execution terminates immediately.

If $X_{\mathrm{cand}}^{(i)}$ is empty, the chain directly proceeds to the next attacker. Otherwise, the candidate is accepted only when it both flips the prediction and satisfies the quality constraints, i.e., $\mathrm{Pert}(X, X_{\mathrm{cand}}^{(i)}) \le \rho$ and $\mathrm{Sim}(X, X_{\mathrm{cand}}^{(i)}) \ge \tau$. If these conditions are met, the chain terminates and returns $X_{\mathrm{cand}}^{(i)}$; otherwise, the candidate is discarded and the next attacker starts a fresh attempt from the same original input $X$ under the remaining budget.

\section{Experiments}
We study the following research questions:
\begin{itemize}[leftmargin=*, noitemsep, topsep=0pt]
   \item RQ1: How does {\OURS} compare with baselines under a fixed query budget?
   \item RQ2: How does the performance of {\OURS} change under different query budgets?
   \item RQ3: How does the performance of {\OURS} change under different perturbation thresholds?
   \item RQ4: Can the fixed global chain selected on the local dataset and local model remain effective across target datasets and target models?
   \item RQ5: Can {\OURS} outperform manually constructed attack chains?
   \item RQ6: How do the key components of {\OURS} affect performance?
\end{itemize}

\subsection{Experimental Setup}
\noindent\textbf{Datasets.} We use five public text classification datasets in total: MR~\cite{pang2005seeing}, SST-2~\cite{socher2013sst}, AG News~\cite{zhang2015character}, Yahoo~\cite{zhang2015character}, and Yelp~\cite{zhang2015character}. MR serves as the local dataset for the global attack chain search, while SST-2, AG News, Yahoo, and Yelp are treated as target datasets.
Table~\ref{tab:dataset} summarizes their scale, label space, and average input length.

\begin{table}[t]
\centering
\small
\begin{tabular}{lcccc}
\toprule
\textbf{Dataset} & \textbf{Train} & \textbf{Test} & \textbf{Class} & \textbf{Length} \\
\midrule
SST-2 & 70K   & 2K   & 2  & 8   \\
MR    & 9K    & 1K   & 2  & 18  \\
Yelp  & 560K  & 38K  & 2  & 133 \\
AG    & 120K  & 7.6K & 4  & 43  \\
Yahoo & 1400K & 60K  & 10 & 151 \\
\bottomrule
\end{tabular}
\caption{Dataset statistics of the five classification benchmarks used in this work.}
\label{tab:dataset}
\end{table}

\medskip
\noindent\textbf{Victim Models.} Following prior hard-label black-box attack studies~\cite{DBLP:conf/aaai/ZhuZ0WL24}, we evaluate {\OURS} on victim models from three architectural paradigms: traditional neural models (WordCNN and WordLSTM), a pre-trained language model (DistilBERT), and large language models (Qwen2.5 in zero-shot and fine-tuned settings). In addition, BERT serves as the local model for the global attack chain search on MR, while the remaining evaluated victims are treated as target models.

\medskip
\noindent\textbf{Baselines.} We compare {\OURS} against nine representative hard-label black-box text attack algorithms: VIWHard~\cite{DBLP:journals/ijon/ZhangWGZWL25}, TextHoaxer~\cite{ye2022texthoaxer}, LeapAttack~\cite{DBLP:conf/kdd/YeCMWM22}, TextHacker~\cite{DBLP:conf/emnlp/YuWC022}, SSPAttack~\cite{DBLP:conf/aaai/0008X0XZMC0Z23}, HQAAttack~\cite{DBLP:journals/corr/abs-2402-01806}, LimeAttack~\cite{DBLP:conf/aaai/ZhuZ0WL24}. In addition, to examine whether a searched chain can also outperform manually designed compositions, we further include two sequential combinations introduced in OpenFact~\cite{DBLP:journals/corr/abs-2409-02649}: BAm2\&Genetic and GSWSE\&T, in the overall comparison.

\begin{itemize}[leftmargin=*, noitemsep, topsep=2pt]
\item \textbf{VIWHard~\cite{DBLP:journals/ijon/ZhangWGZWL25}}: Introduces an important-word discriminator trained on a local surrogate model to identify vulnerable tokens without querying the target. It generates context-aware substitutions via a Masked Language Model (MLM) and optimizes the attack using a genetic algorithm.
\item \textbf{TextHoaxer~\cite{ye2022texthoaxer}}: A greedy heuristic that prioritizes token positions, using a few probing queries to determine whether a substitution is worthwhile before committing.
\item \textbf{LeapAttack~\cite{DBLP:conf/kdd/YeCMWM22}}: Employs finite-difference boundary exploration with gradient-like directions to jointly select positions and synonyms, pushing inputs into the misclassification region.
\item \textbf{TextHacker~\cite{DBLP:conf/emnlp/YuWC022}}: Maintains an online-updated word-importance table derived from edit-flip history, which guides a hybrid local search.
\item \textbf{SSPAttack~\cite{DBLP:conf/aaai/0008X0XZMC0Z23}}: First obtains a successful adversarial example via synonym substitutions, and then reduces perturbations by reverting unnecessary replacements while preserving the attack. This process aims to improve text quality under a limited query budget.
\item \textbf{HQAAttack~\cite{DBLP:journals/corr/abs-2402-01806}}: Focuses on generating high-quality adversarial texts under hard-label feedback. Starting from an adversarial example, it refines substitutions and optimizes critical token positions to reduce perturbation while maintaining attack success.
\item \textbf{LimeAttack~\cite{DBLP:conf/aaai/ZhuZ0WL24}}: Leverages a local explainable method to approximate word importance ranking, and then adopts beam search to find the optimal solution.
\item \textbf{BAm2\&Genetic~\cite{DBLP:journals/corr/abs-2409-02649}}: A combination introduced in OpenFact, where the Genetic Algorithm is applied for cases in which BAm2 fails, in order to bolster the attack success rate.
\item \textbf{GSWSE\&TF~\cite{DBLP:journals/corr/abs-2409-02649}}: A method introduced in OpenFact, where TextFooler is applied to texts for which GSWSE is unable to change the model decision.
\end{itemize}

\medskip
\noindent\textbf{Evaluation Metrics.} Consistent with prior work such as LimeAttack and VIWHard, we use Attack Success Rate (ASR), Perturbation Rate (Pert), and Semantic Similarity (Sim) as evaluation metrics. ASR is computed over originally correct samples only, and measures the proportion of valid attack targets that are successfully flipped by the attack. To assess adversarial text quality, Pert quantifies the fraction of modified tokens, while Sim measures the degree of semantic preservation with respect to the original input using the same sentence-level similarity module as the attack constraint.

\medskip
\noindent\textbf{Implementation Details.} A key protocol of our evaluation is that the attack chain is searched only once on the local dataset under the local model and then reused across all target datasets and target models. Concretely, each candidate chain is an ordered length-$L=3$ sequence sampled without repetition from the seven hard-label attackers. NSGA-II optimizes these candidates under a dual objective of maximizing ASR and minimizing Pert on $N=100$ correctly classified local samples, using the same constraints as the final evaluation setting ($B=1000$, $\rho=0.1$, $\tau=0.9$). TOPSIS is then applied to select a single fixed global chain for all target evaluations:
\[
  \textbf{HQAAttack} \;\to\; \textbf{VIWHard} \;\to\; \textbf{TextHacker}.
\]
For the final attack setting, we use $B=1000$, $M=50$, $\rho=0.1$, and $\tau=0.9$. For NSGA-II, we use population size $P=30$, maximum generations $G=10$, crossover probability $p_c=0.85$, and mutation probability $p_m=0.2$. Each target evaluation uses $N=1000$ samples. All results are averaged over three independent runs on an NVIDIA RTX 3080 Ti GPU (13 GB).

\subsection{Overall Result (RQ1)}
Table~\ref{tab:overall} shows that the searched chain consistently achieves higher ASR than strong standalone baselines while keeping Pert in a comparable range on both conventional classifiers and LLM victims. On WordCNN for SST-2, {\OURS} reaches 52.3\% ASR, clearly above LeapAttack and HQAAttack (both 36.5\%), with similar Pert (6.4\%). The same trend appears on Qwen2.5-FT for SST-2, where {\OURS} achieves 42.0\% ASR, outperforming HQAAttack (37.5\%) and LeapAttack (35.4\%) with comparable perturbation (6.8\%). Moreover, {\OURS}'s results consistently surpass those of the two combination baselines adapted from OpenFact.

This gain is better explained by attacker complementarity. Different attackers follow different search trajectories and therefore succeed on different subsets of samples.  The attack chain exploits this complementarity, allowing later attackers to cover residual cases missed by earlier ones and thus improve overall ASR. In addition, {\OURS} outperforms the combination baselines because OpenFact uses manually constructed combinations, whereas our method searches for the attack chain automatically, allowing it to better exploit the complementarity among different attackers.

\begin{table}[htbp!]
\centering
\scriptsize
\setlength{\tabcolsep}{2.5pt}
\renewcommand{\arraystretch}{0.95}
\resizebox{\textwidth}{!}{
\begin{tabular}{ll cc cc cc cc}
\toprule
\multirow{2}{*}{\textbf{Model}} & \multirow{2}{*}{\textbf{Method}}
& \multicolumn{2}{c}{\textbf{SST-2}}
& \multicolumn{2}{c}{\textbf{AG}}
& \multicolumn{2}{c}{\textbf{Yahoo}}
& \multicolumn{2}{c}{\textbf{Yelp}} \\
\cmidrule(lr){3-4} \cmidrule(lr){5-6} \cmidrule(lr){7-8} \cmidrule(lr){9-10}
& & \textbf{ASR}$\uparrow$ & \textbf{Pert}$\downarrow$
  & \textbf{ASR}$\uparrow$ & \textbf{Pert}$\downarrow$
  & \textbf{ASR}$\uparrow$ & \textbf{Pert}$\downarrow$
  & \textbf{ASR}$\uparrow$ & \textbf{Pert}$\downarrow$ \\
\midrule

\multirow{10}{*}{\textbf{WordCNN}}
& \textbf{\OURS}      & \textbf{52.3} & 6.4 & \textbf{60.3} & 6.0 & \textbf{79.8} & 4.2 & \textbf{85.2} & 3.7 \\
& LimeAttack          & 35.3 & 6.7 & 26.7 & 3.1 & 43.6 & 2.9 & 31.4 & \textbf{2.4} \\
& VIWHard             & 32.9 & 6.6 & 18.9 & \textbf{2.6} & 33.8 & 2.4 & 18.6 & \textbf{2.4} \\
& TextHacker          & 22.4 & 7.0 & 43.3 & 6.0 & 69.0 & 4.6 & 65.1 & 4.6 \\
& TextHoaxer          & 34.1 & 6.4 & 53.3 & 6.1 & 66.2 & 4.6 & 74.4 & 4.0 \\
& LeapAttack          & 36.5 & \textbf{6.2} & 59.1 & 5.5 & 56.3 & 3.3 & 54.7 & 3.4 \\
& HQAAttack           & 36.5 & 6.3 & 58.9 & 5.7 & 76.1 & 4.2 & 81.4 & 3.7 \\
& SSPAttack           & 30.6 & 6.4 & 45.6 & 5.7 & 64.8 & \textbf{1.2} & 65.1 & 4.2 \\
& BAM2\&Genetic       & 33.7 & 6.8 & 30.5 & 5.9 & 41.9 & 3.6 & 47.8 & 3.5 \\
& GSWSE\&TF           & 48.2 & 6.9 & 28.8 & 4.8 & 35.2 & 3.1 & 33.5 & 3.3 \\
\midrule

\multirow{10}{*}{\textbf{WordLSTM}}
& \textbf{\OURS}      & \textbf{38.7} & 7.0 & \textbf{42.5} & 6.0 & \textbf{66.1} & 4.2 & \textbf{89.3} & 3.7 \\
& LimeAttack          & 28.6 & 6.9 & 26.4 & 3.8 & 34.8 & 3.0 & 29.1 & 2.7 \\
& VIWHard             & 26.1 & 6.9 & 25.7 & \textbf{2.7} & 32.4 & \textbf{2.3} & 23.3 & 2.8 \\
& TextHacker          & 21.4 & 6.7 & 32.3 & 5.8 & 52.9 & 4.4 & 73.3 & 4.2 \\
& TextHoaxer          & 25.6 & \textbf{6.3} & 36.6 & 5.6 & 62.2 & 4.2 & 79.1 & \textbf{1.0} \\
& LeapAttack          & 29.6 & 6.4 & 36.2 & 5.6 & 51.5 & 3.7 & 70.9 & 3.7 \\
& HQAAttack           & 28.4 & 7.0 & 42.2 & 5.9 & 61.7 & 3.8 & 81.4 & \textbf{1.0} \\
& SSPAttack           & 24.9 & 7.2 & 36.4 & 5.6 & 61.2 & 4.4 & 68.6 & 3.7 \\
& BAM2\&Genetic       & 27.8 & 6.9 & 31.5 & 5.8 & 44.3 & 3.9 & 62.7 & 3.6 \\
& GSWSE\&TF           & 35.2 & 6.6 & 34.7 & 5.7 & 50.8 & 3.6 & 74.5 & 3.5 \\
\midrule

\multirow{10}{*}{\textbf{DistilBERT}}
& \textbf{\OURS}      & \textbf{44.8} & 6.9 & \textbf{36.8} & 6.0 & \textbf{61.2} & 3.9 & \textbf{75.1} & 4.0 \\
& LimeAttack          & 31.6 & 6.8 & 12.9 & 4.2 & 30.6 & 2.8 & 19.0 & 2.4 \\
& VIWHard             & 25.8 & 6.3 & 12.2 & \textbf{2.8} & 27.7 & \textbf{2.4} & 12.5 & \textbf{1.9} \\
& TextHacker          & 15.1 & \textbf{6.2} & 9.8 & 4.8 & 21.8 & 3.2 & 19.0 & 3.0 \\
& TextHoaxer          & 36.6 & 6.7 & 27.3 & 5.1 & 45.8 & 3.6 & 63.8 & 4.1 \\
& LeapAttack          & 18.7 & 6.9 & 15.4 & 5.7 & 36.8 & 4.0 & 52.0 & 4.3 \\
& HQAAttack           & 43.8 & 7.0 & 32.5 & 6.0 & 55.6 & 4.0 & 73.9 & 3.9 \\
& SSPAttack           & 36.0 & 7.0 & 23.8 & 5.6 & 46.9 & 3.9 & 58.2 & 4.1 \\
& BAM2\&Genetic       & 24.4 & 7.6 & 28.4 & 3.4 & 28.5 & 4.8 & 40.7 & 3.5 \\
& GSWSE\&TF           & 42.1 & 6.8 & 36.3 & 3.0 & 17.4 & 3.4 & 18.4 & 2.9 \\
\midrule

\multirow{10}{*}{\textbf{\shortstack{\textbf{Qwen2.5}\\\textbf{(Zero-shot)}}}}
& \textbf{\OURS}      & \textbf{48.7} & 6.3 & \textbf{60.4} & 2.8 & \textbf{51.5} & 2.2 & \textbf{40.1} & 2.1 \\
& LimeAttack          & 47.2 & 6.3 & 53.6 & 3.4 & 49.4 & 2.3 & 38.3 & 2.2 \\
& VIWHard             & 45.4 & 6.7 & 45.4 & \textbf{2.3} & 38.5 & \textbf{1.4} & 36.1 & \textbf{1.7} \\
& TextHacker          & 17.1 & \textbf{6.2} & 19.5 & 4.6 & 14.2 & 4.7 & 13.4 & 3.5 \\
& TextHoaxer          & 11.6 & 7.0 & 10.7 & 3.9 & 15.6 & 3.4 & 9.5 & 2.3 \\
& LeapAttack          & 34.6 & \textbf{6.2} & 31.6 & 4.1 & 33.6 & 3.1 & 32.2 & 2.3 \\
& HQAAttack           & 18.5 & 6.7 & 19.3 & 4.3 & 18.7 & 3.5 & 16.1 & 2.8 \\
& SSPAttack           & 10.6 & 6.4 & 12.4 & 3.9 & 15.8 & 3.1 & 14.7 & 2.2 \\
& BAM2\&Genetic       & 29.8 & 6.4 & 27.9 & 4.2 & 26.7 & 3.4 & 24.6 & 2.6 \\
& GSWSE\&TF           & 31.5 & 6.5 & 29.4 & 4.0 & 28.9 & 3.0 & 26.3 & 2.4 \\
\midrule

\multirow{10}{*}{\shortstack{\textbf{Qwen2.5}\\\textbf{(Fine-tuned)}}}
& \textbf{\OURS}      & \textbf{42.0} & 6.8 & \textbf{41.2} & 6.0 & \textbf{56.5} & 4.4 & \textbf{64.2} & 4.7 \\
& LimeAttack          & 22.9 & \textbf{6.4} & 15.7 & 3.4 & 22.1 & \textbf{2.2} & 16.2 & 3.5 \\
& VIWHard             & 17.7 & 7.0 & 14.6 & \textbf{2.8} & 20.7 & 3.0 & 10.4 & \textbf{2.3} \\
& TextHacker          & 18.7 & 7.4 & 15.7 & 6.2 & 23.3 & 4.3 & 32.0 & 3.7 \\
& TextHoaxer          & 29.2 & 6.7 & 30.3 & 3.7 & 36.3 & 2.3 & 42.4 & 4.8 \\
& LeapAttack          & 35.4 & 6.7 & 40.1 & 3.8 & 46.7 & 2.4 & 58.8 & 4.9 \\
& HQAAttack           & 37.5 & 6.9 & 40.4 & 4.0 & 55.8 & 2.4 & 61.3 & 4.8 \\
& SSPAttack           & 23.9 & 6.9 & 17.9 & 3.7 & 19.4 & 2.3 & 38.5 & 4.6 \\
& BAM2\&Genetic       & 33.8 & 6.9 & 34.6 & 5.8 & 46.1 & 3.9 & 51.7 & 4.5 \\
& GSWSE\&TF           & 41.6 & 6.7 & 37.1 & 6.3 & 42.3 & 3.8 & 56.4 & 4.6 \\

\bottomrule
\end{tabular}
}
\caption{Comprehensive performance comparison (ASR \% and Pert \%) across all victim models and datasets under a 1000-Query Budget.}
\label{tab:overall}
\end{table}

\subsection{ASR under Varying Query Budgets (RQ2)}
\begin{figure}[t]
  \centering
  \begin{subfigure}{0.49\linewidth}
    \centering
    \includegraphics[width=\linewidth]{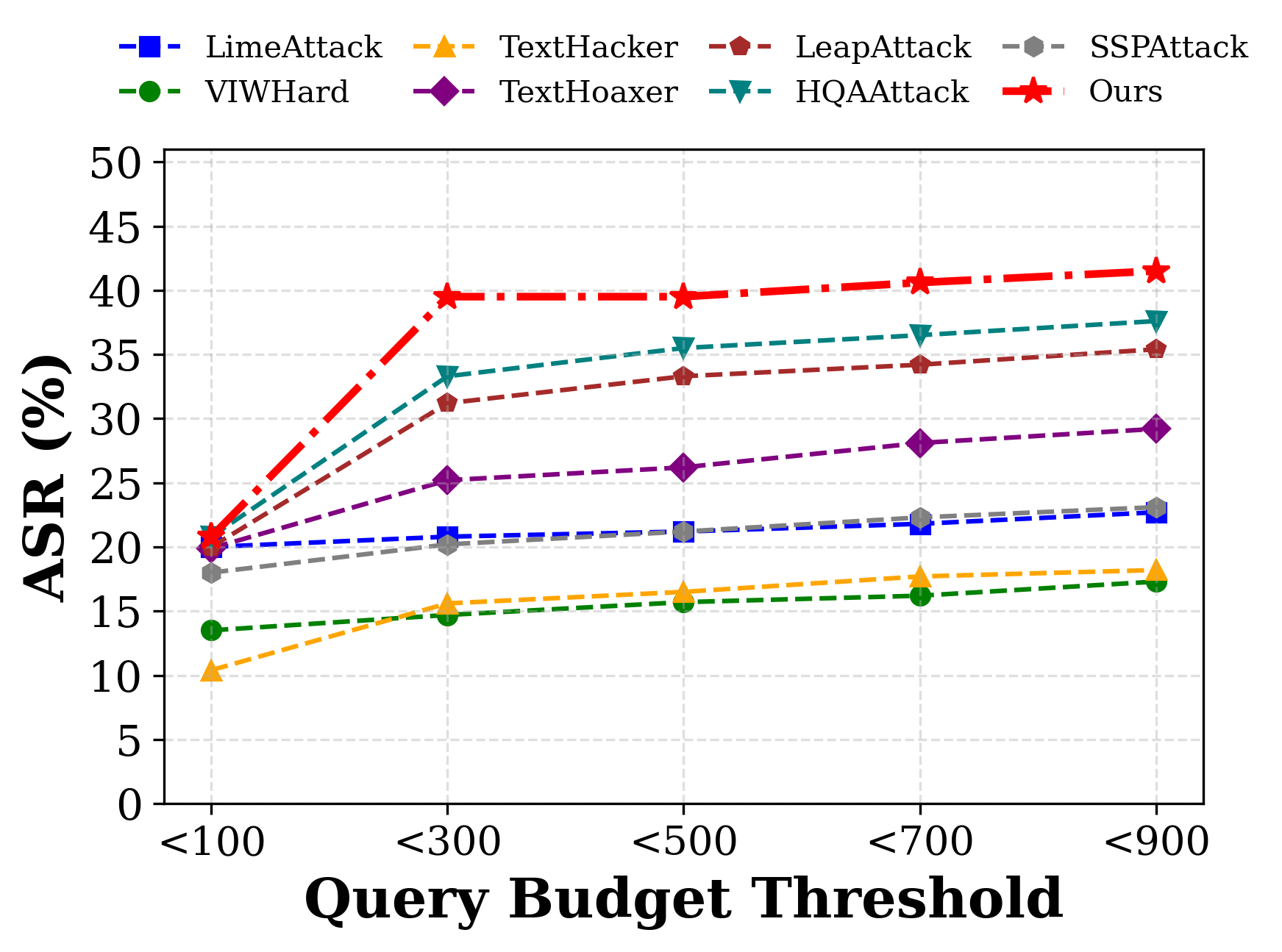}
    \caption{SST-2}
    \label{fig:query-sst2}
  \end{subfigure}\hfill
  \begin{subfigure}{0.49\linewidth}
    \centering
    \includegraphics[width=\linewidth]{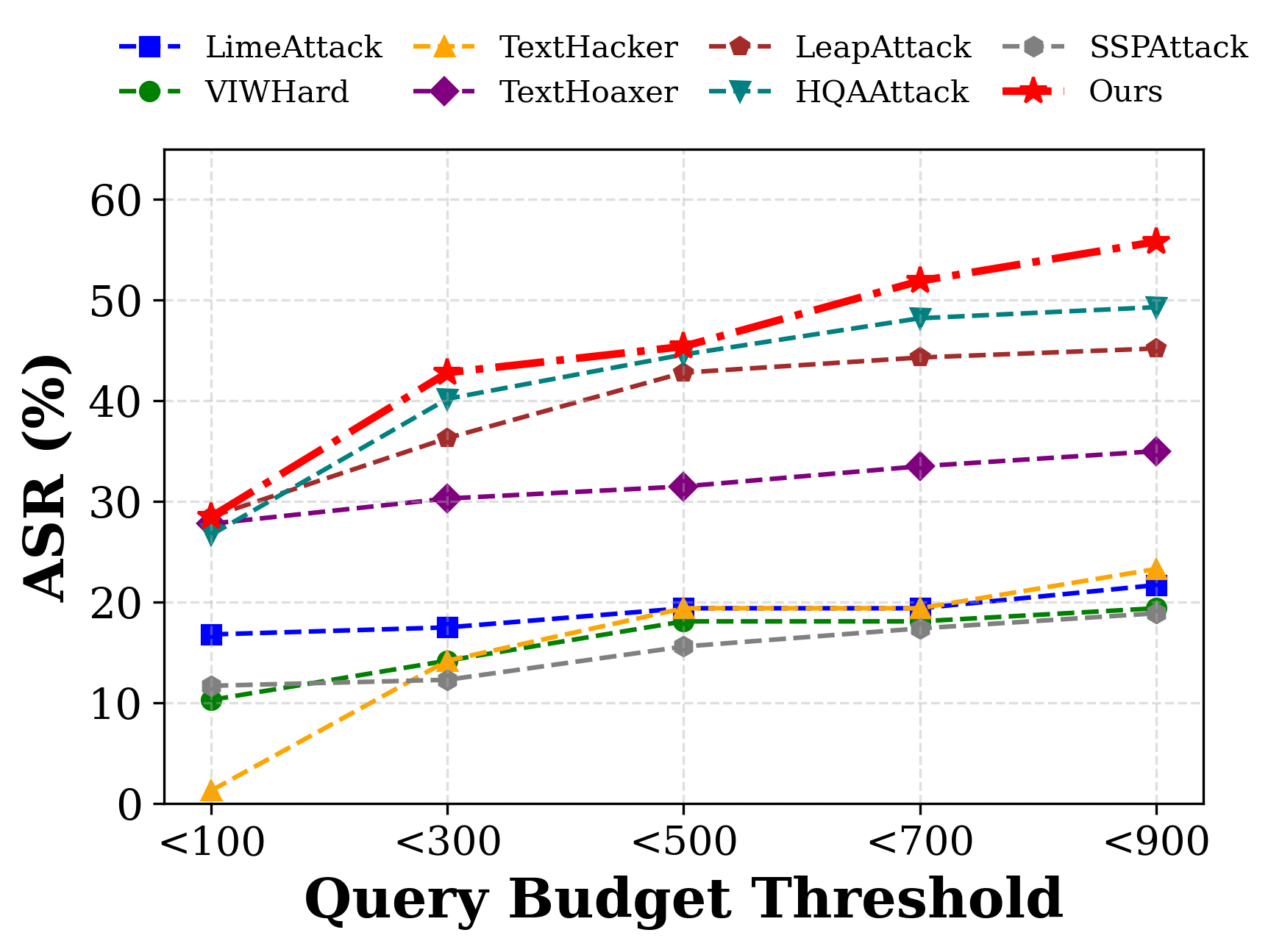}
    \caption{Yahoo}
    \label{fig:query-yahoo}
  \end{subfigure}
  \caption{ASR under different query budget thresholds.}
  \label{fig:query-threshold}
\end{figure}

Figure~\ref{fig:query-threshold} shows the ASR trends on Qwen2.5-FT for SST-2 and Yahoo under five query budgets, with $B \in \{100, 300, 500, 700, 900\}$. Across all budgets, {\OURS} stays above the baselines, and the gap becomes more visible once the budget is large enough for later attackers to act on residual hard examples, especially on Yahoo. This behavior is consistent with the chain acting as a fallback over complementary search trajectories: later attackers do not receive extra budget, but they can still solve cases that earlier search patterns miss. The improvement therefore comes from broader sample coverage, rather than from budget inflation.

\subsection{ASR under Varying PERT (RQ3)}
\begin{figure*}[t]
  \centering
  \begin{subfigure}{0.48\textwidth}
    \centering
    \includegraphics[width=\linewidth]{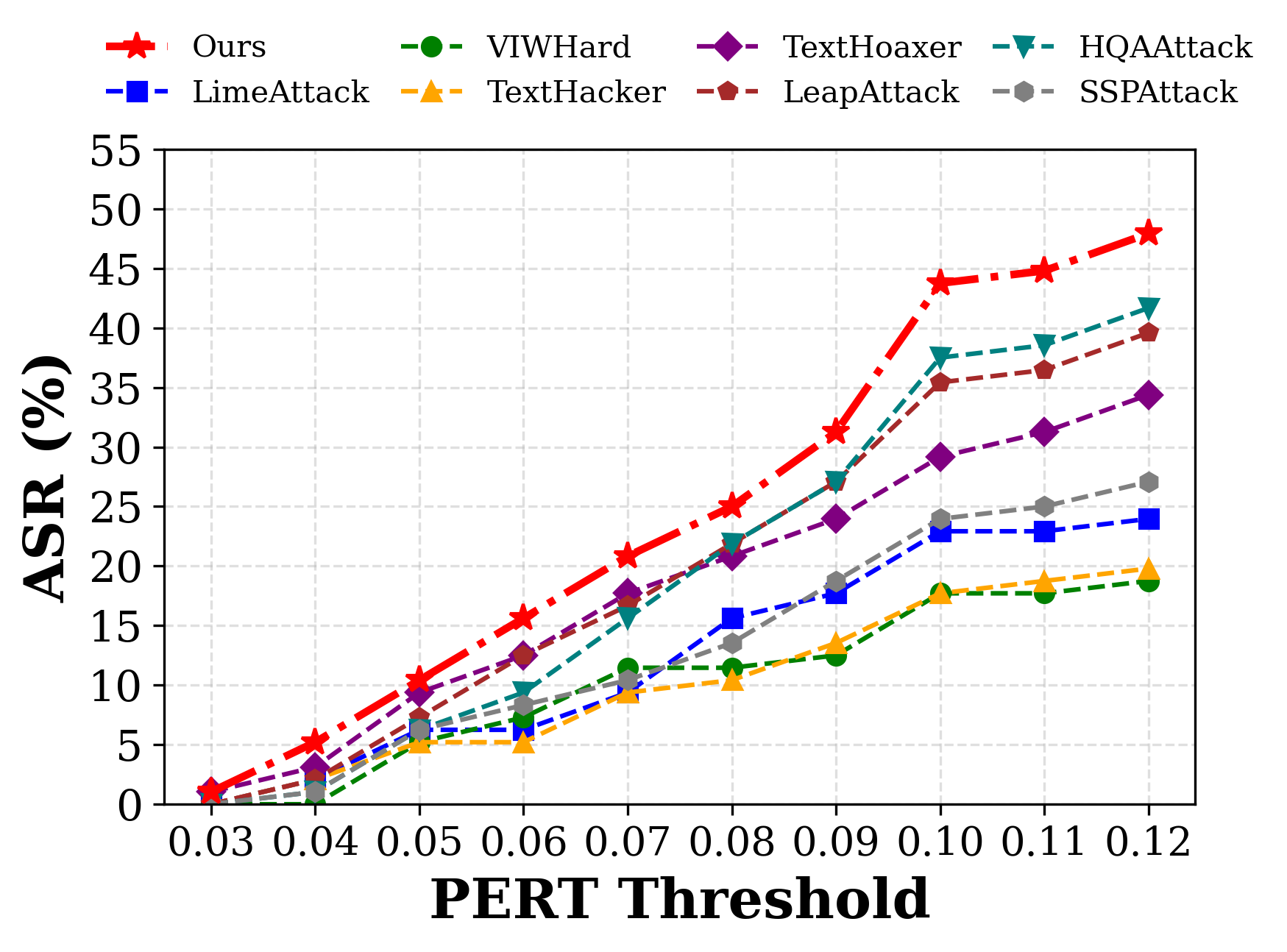}
    \caption{SST-2 (Qwen2.5-FT)}
    \label{fig:pert-sst2-ft}
  \end{subfigure}\hfill
  \begin{subfigure}{0.48\textwidth}
    \centering
    \includegraphics[width=\linewidth]{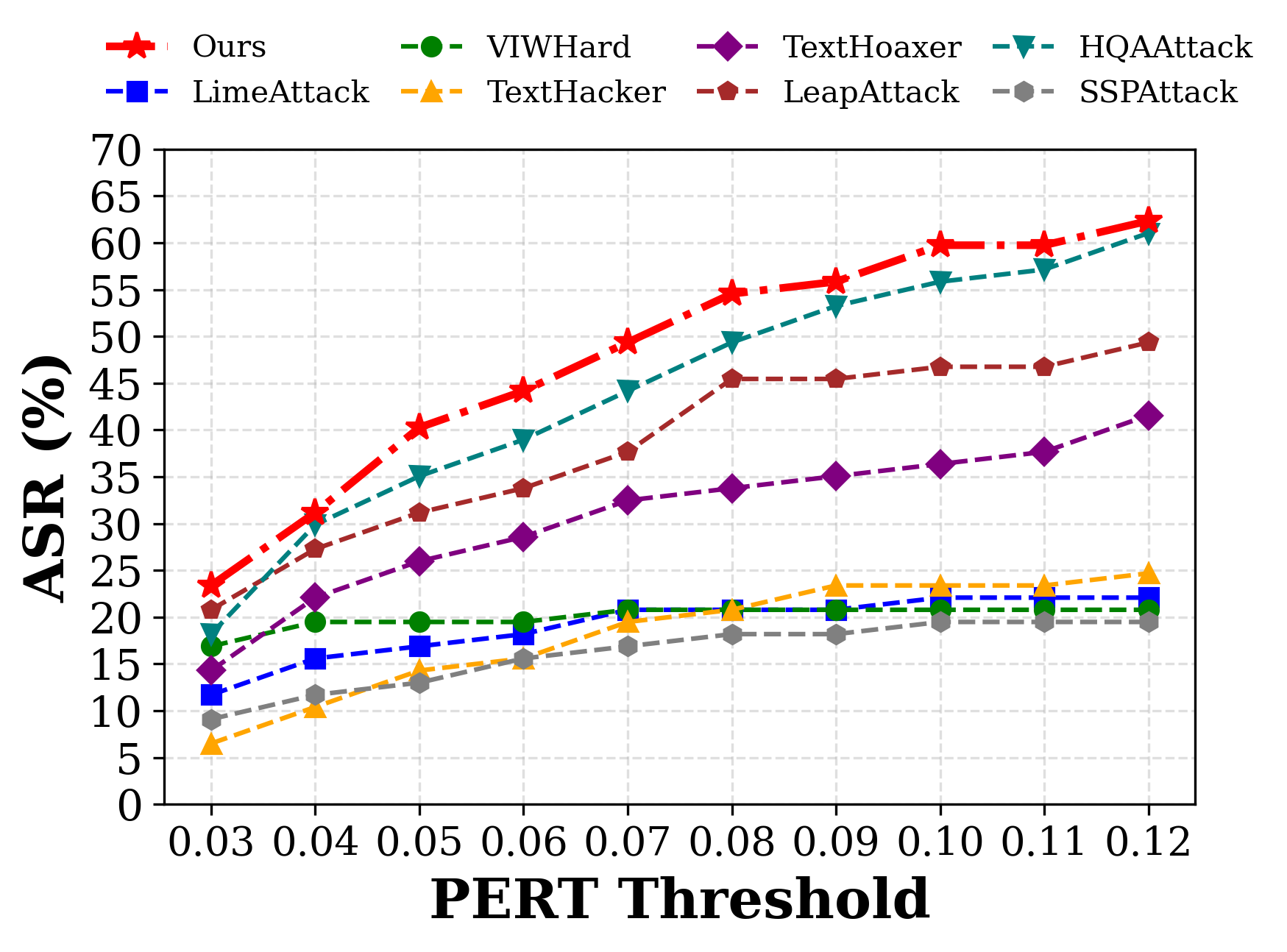}
    \caption{Yahoo (Qwen2.5-FT)}
    \label{fig:pert-yahoo-ft}
  \end{subfigure}

  \medskip

  \begin{subfigure}{0.48\textwidth}
    \centering
    \includegraphics[width=\linewidth]{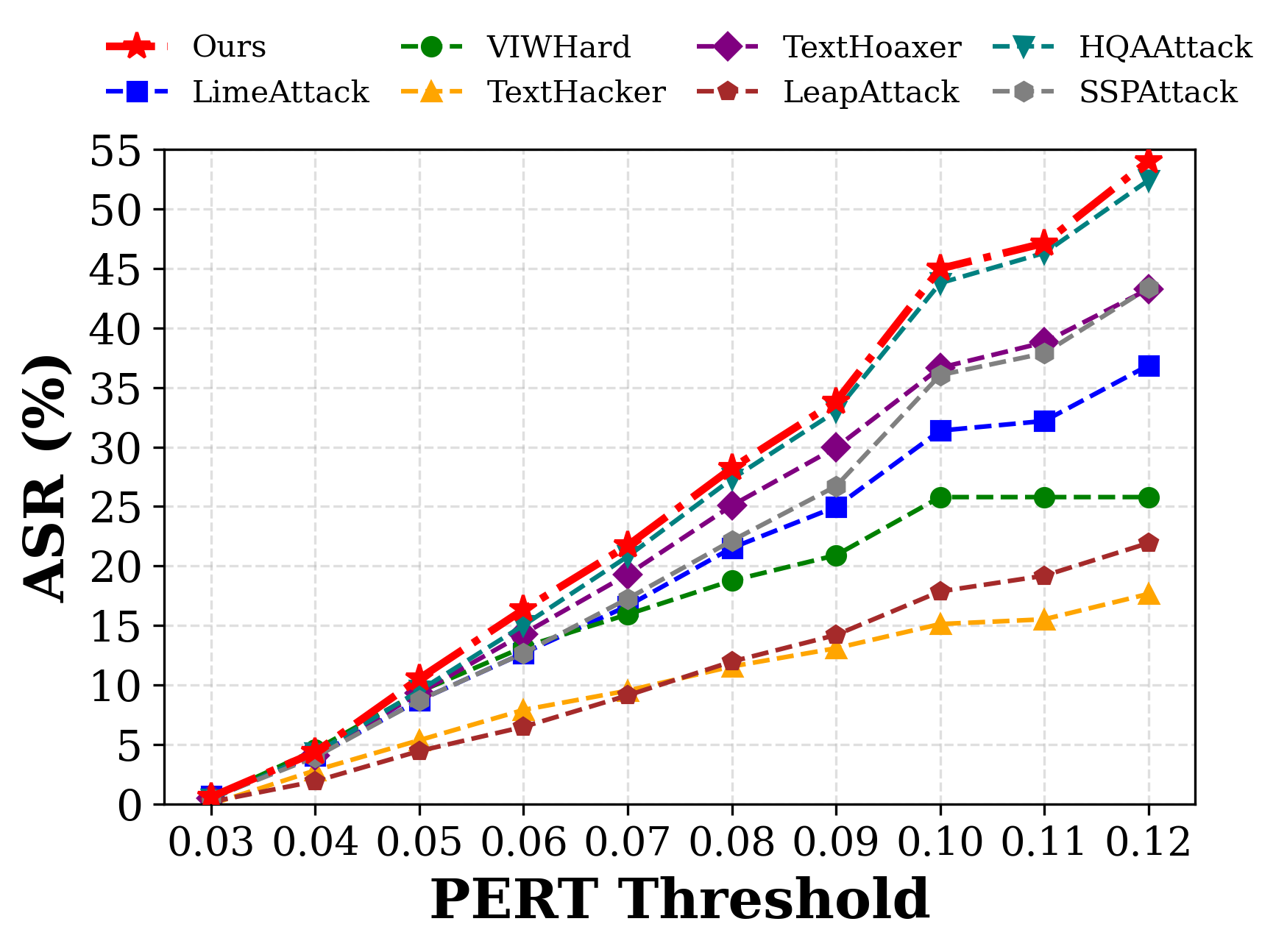}
    \caption{SST-2 (DistilBERT)}
  \end{subfigure}\hfill
  \begin{subfigure}{0.48\textwidth}
    \centering
    \includegraphics[width=\linewidth]{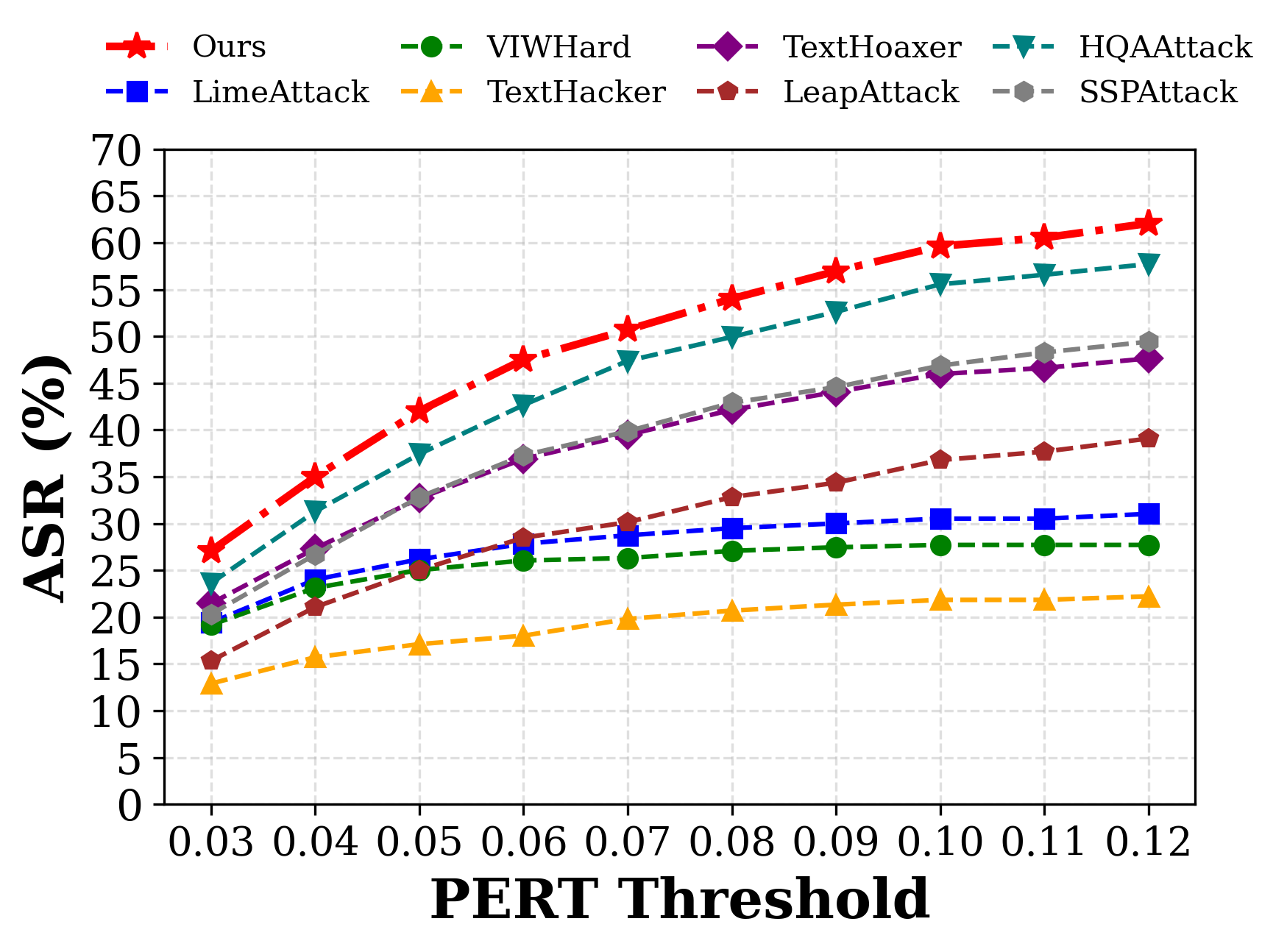}
    \caption{Yahoo (DistilBERT)}
  \end{subfigure}
  \caption{ASR--PERT curves under different perturbation thresholds on Qwen2.5-FT and DistilBERT. The first row shows Qwen2.5-FT, and the second row shows DistilBERT.}
  \label{fig:pert-threshold}
\end{figure*}

Figure~\ref{fig:pert-threshold} reports four ASR--PERT curves on Qwen2.5-FT and DistilBERT for SST-2 and Yahoo under perturbation thresholds $\rho \in [0.03, 0.12]$. The clearest pattern appears on Yahoo, where {\OURS} traces the upper frontier across almost the entire threshold range for both victim models; on SST-2, the curves are closer, but {\OURS} still stays on or near the favorable frontier. This shows that its gains do not come from relaxing the perturbation constraint. This stability is consistent with the design of {\OURS}: NSGA-II jointly optimizes ASR and Pert, and TOPSIS selects a balanced chain rather than one that attains high ASR through larger edits. As a result, the selected chain shows a smaller ASR drop as the perturbation threshold becomes stricter, while later attackers can still cover some valid low-perturbation adversarial examples missed by earlier ones.

\subsection{Transferability of the Fixed Global Chain (RQ4)}
\begin{figure*}[t]
  \centering
  \begin{subfigure}{0.24\textwidth}
    \centering
    \includegraphics[width=\linewidth]{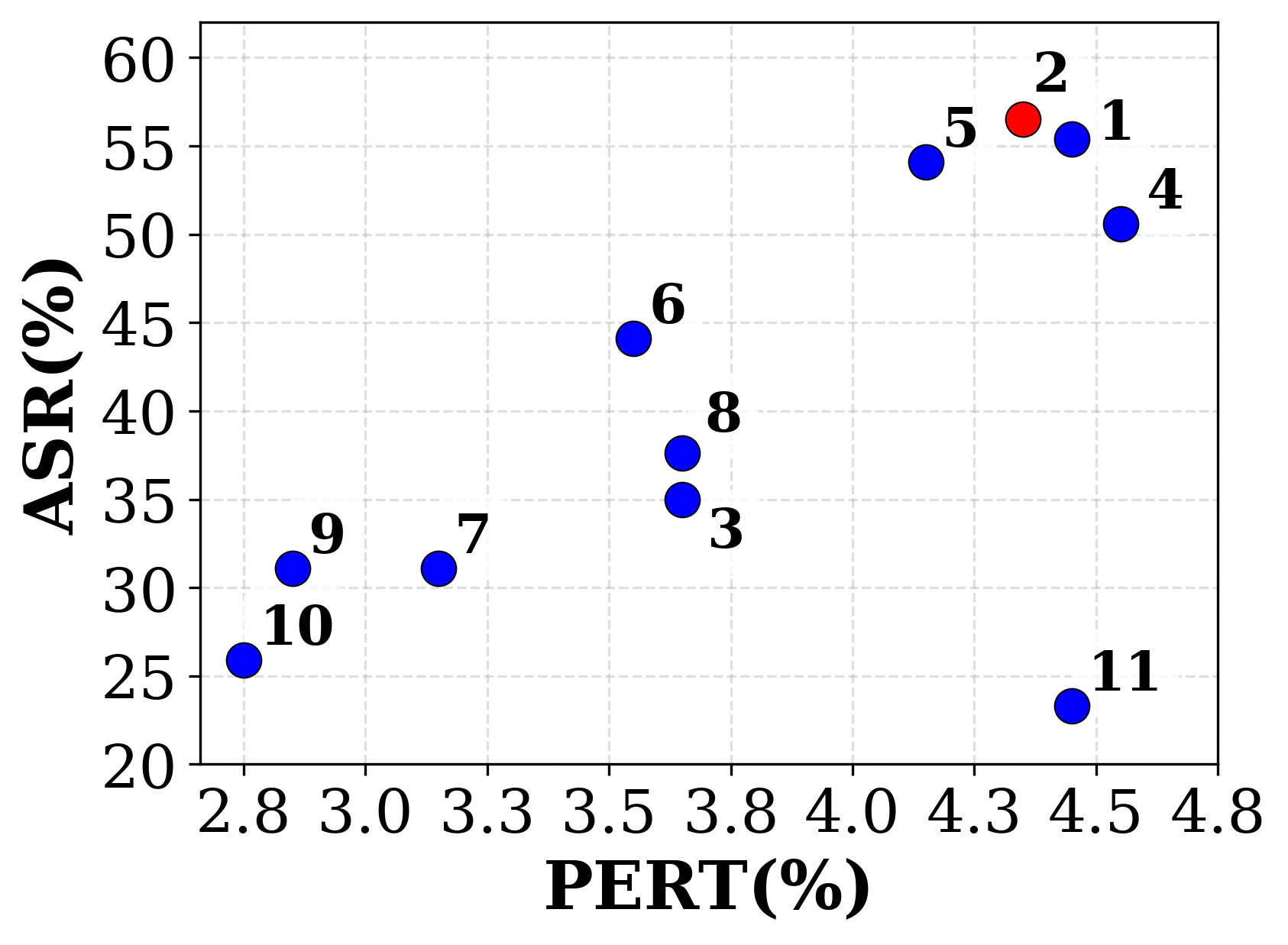}
    \caption{Yahoo (Qwen2.5-FT)}
    \label{fig:topsis-qw-yahoo}
  \end{subfigure}\hfill
  \begin{subfigure}{0.24\textwidth}
    \centering
    \includegraphics[width=\linewidth]{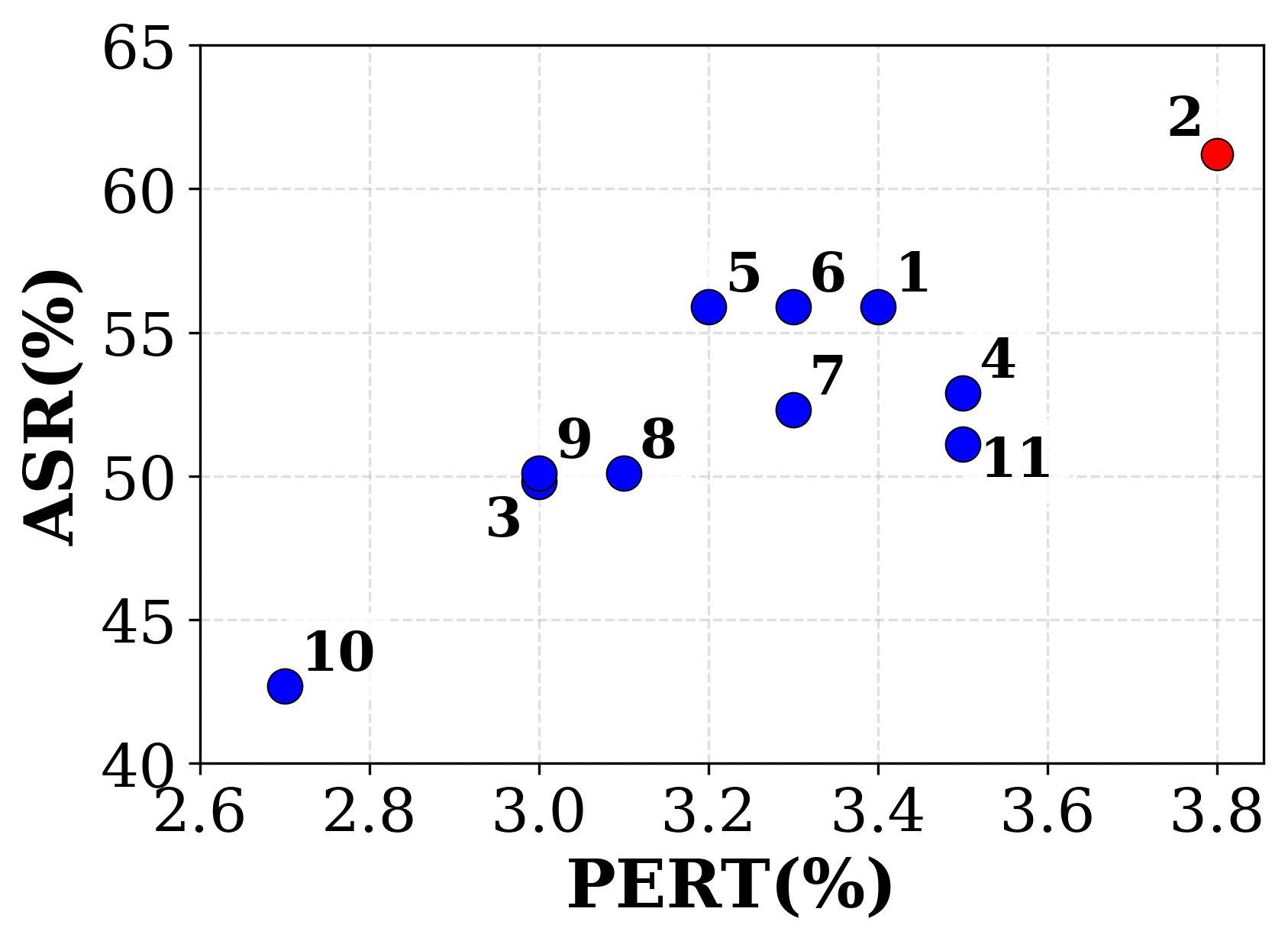}
    \caption{Yahoo (DistilBERT)}
    \label{fig:topsis-dis-yahoo}
  \end{subfigure}\hfill
  \begin{subfigure}{0.24\textwidth}
    \centering
    \includegraphics[width=\linewidth]{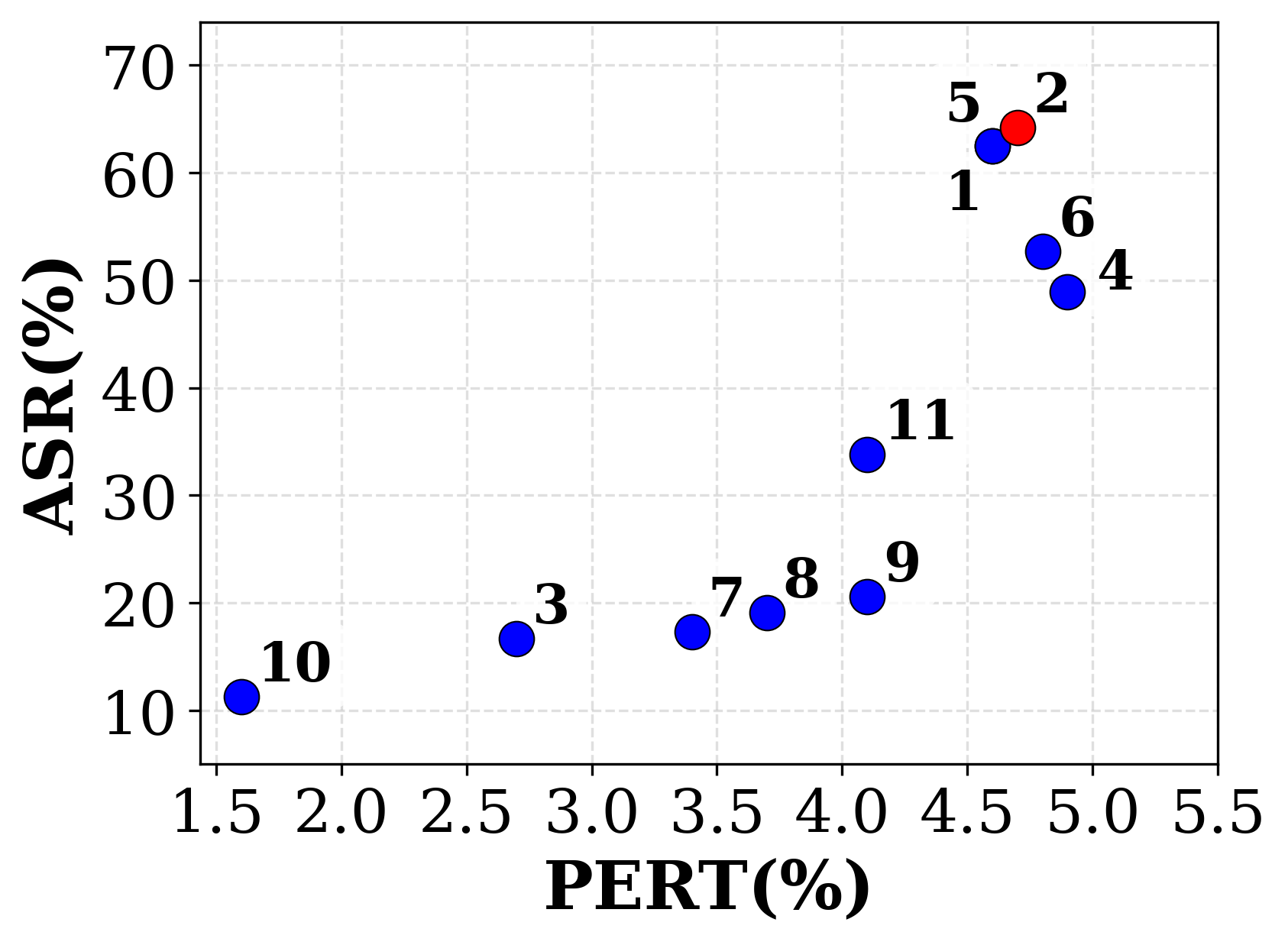}
    \caption{Yelp (Qwen2.5-FT)}
  \end{subfigure}\hfill
  \begin{subfigure}{0.24\textwidth}
    \centering
    \includegraphics[width=\linewidth]{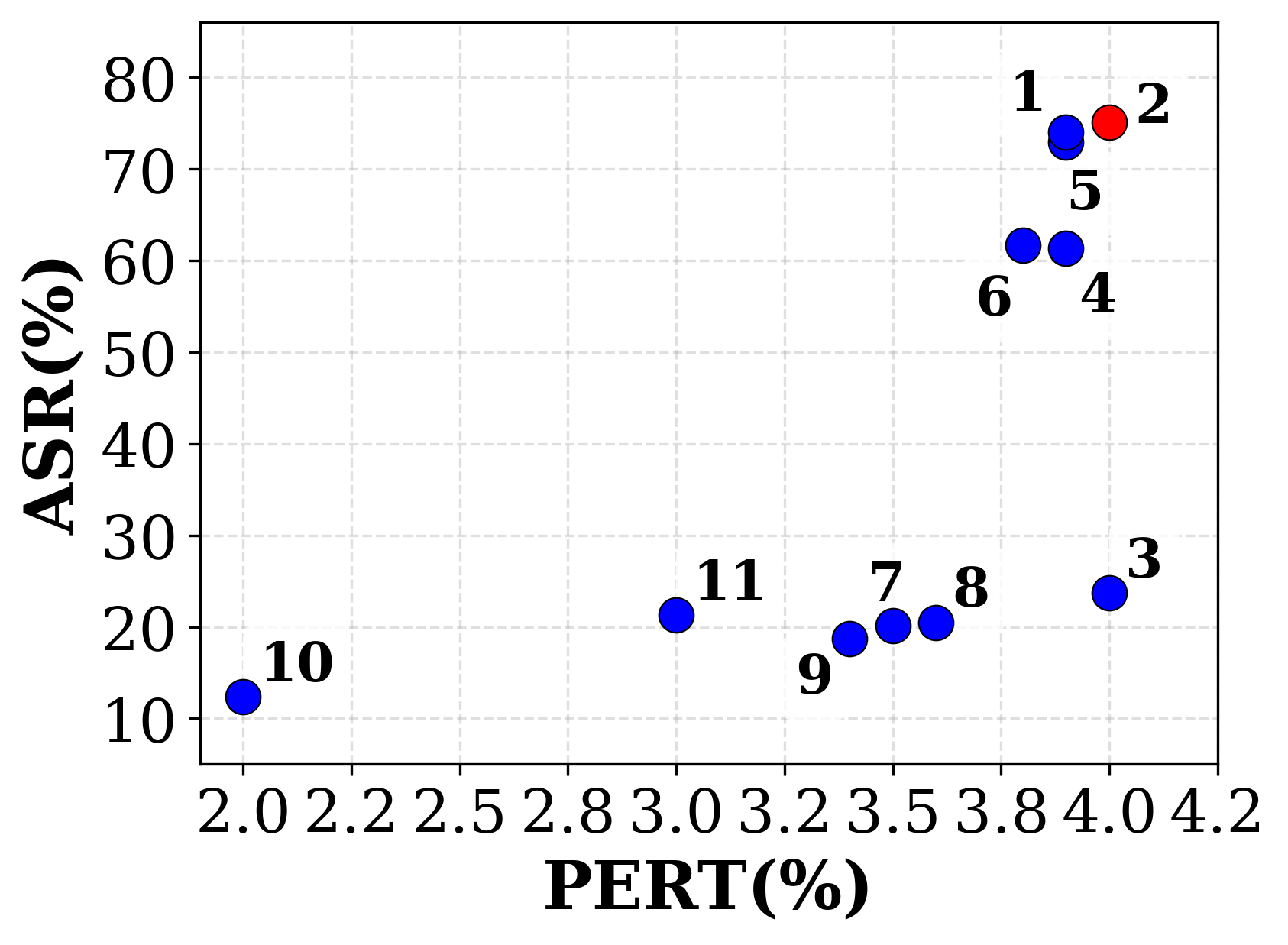}
    \caption{Yelp (DistilBERT)}
  \end{subfigure}

  \medskip

  \begin{subfigure}{0.24\textwidth}
    \centering
    \includegraphics[width=\linewidth]{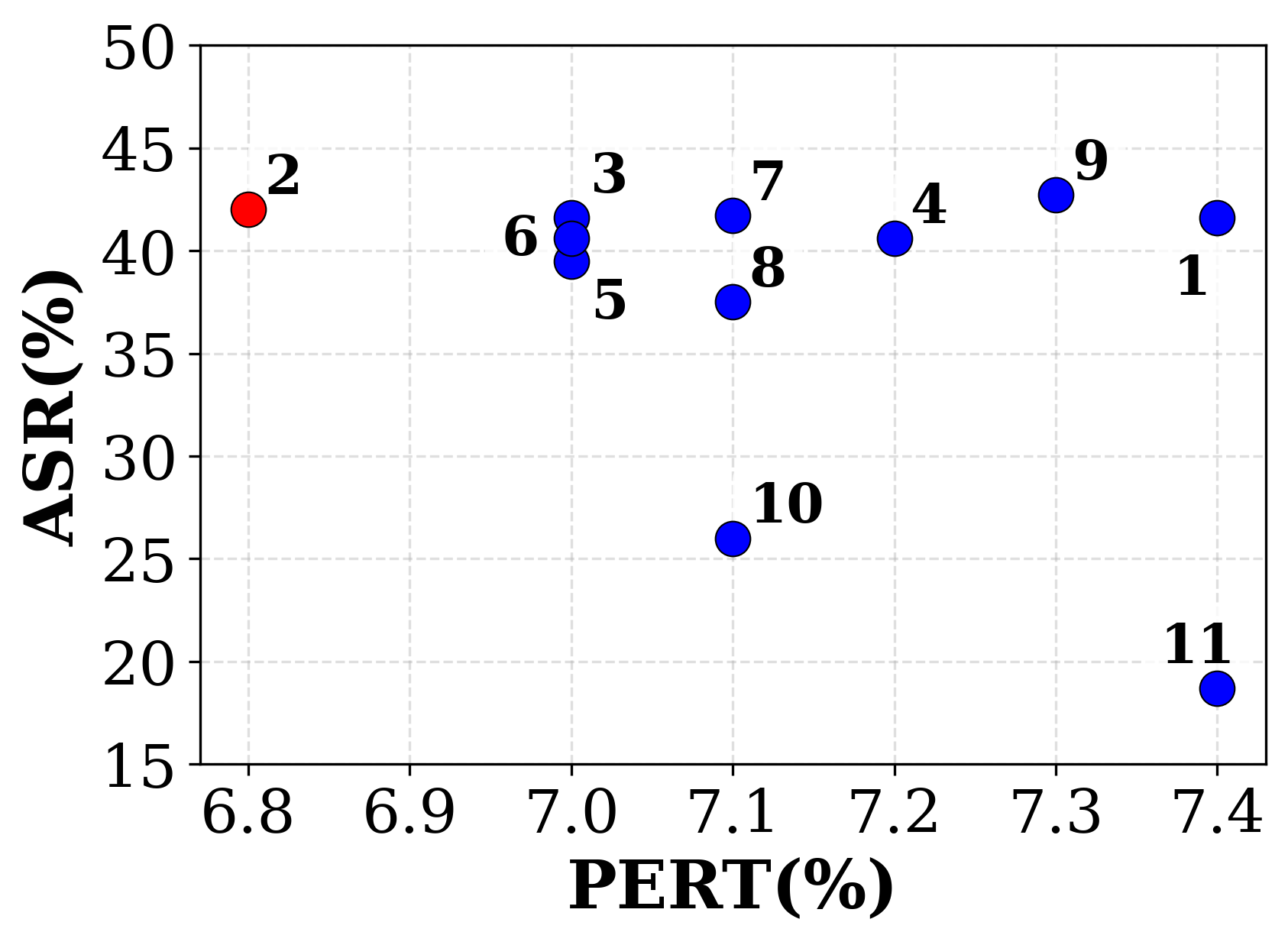}
    \caption{SST-2 (Qwen2.5-FT)}
  \end{subfigure}\hfill
  \begin{subfigure}{0.24\textwidth}
    \centering
    \includegraphics[width=\linewidth]{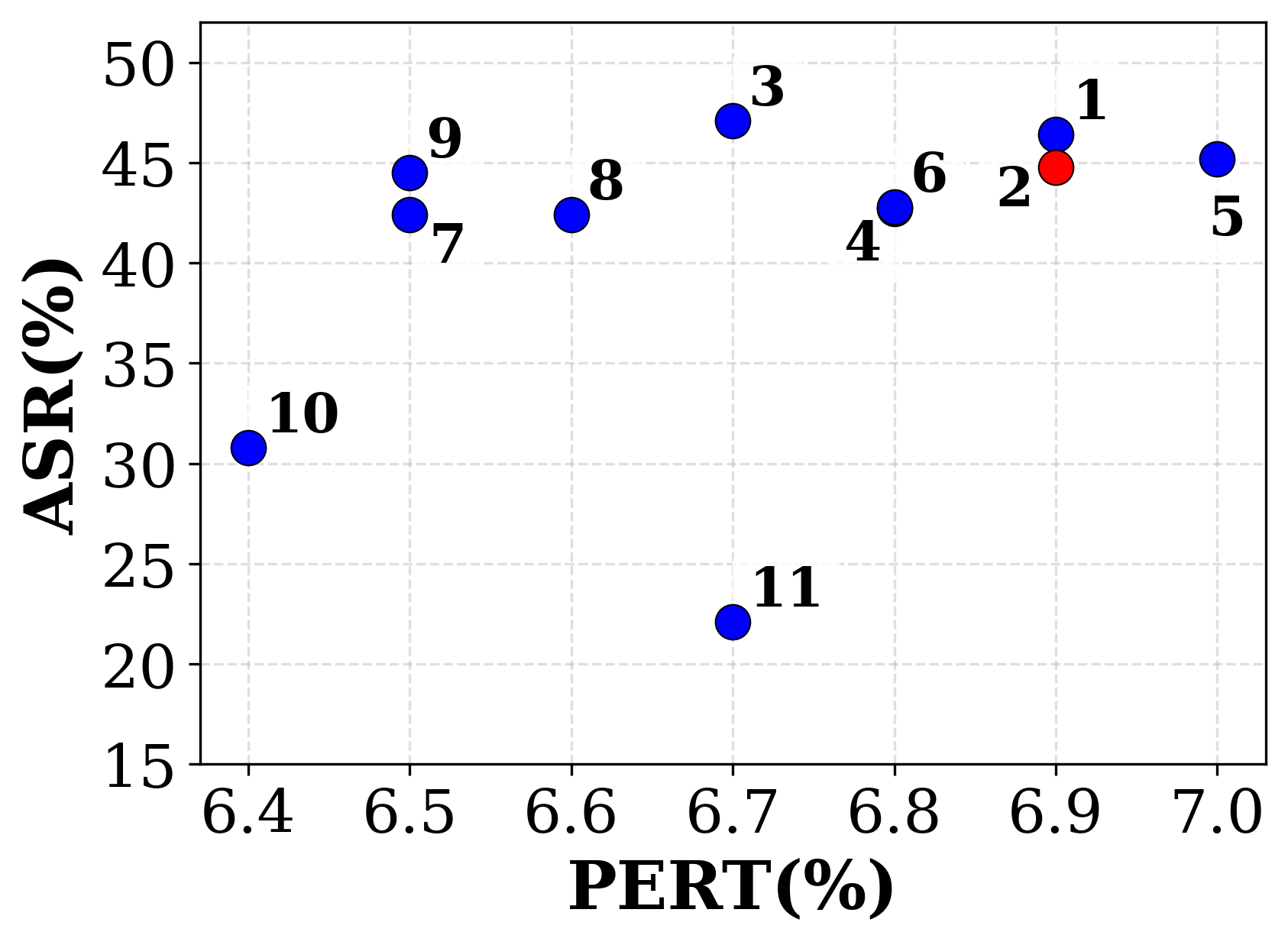}
    \caption{SST-2 (DistilBERT)}
  \end{subfigure}\hfill
  \begin{subfigure}{0.24\textwidth}
    \centering
    \includegraphics[width=\linewidth]{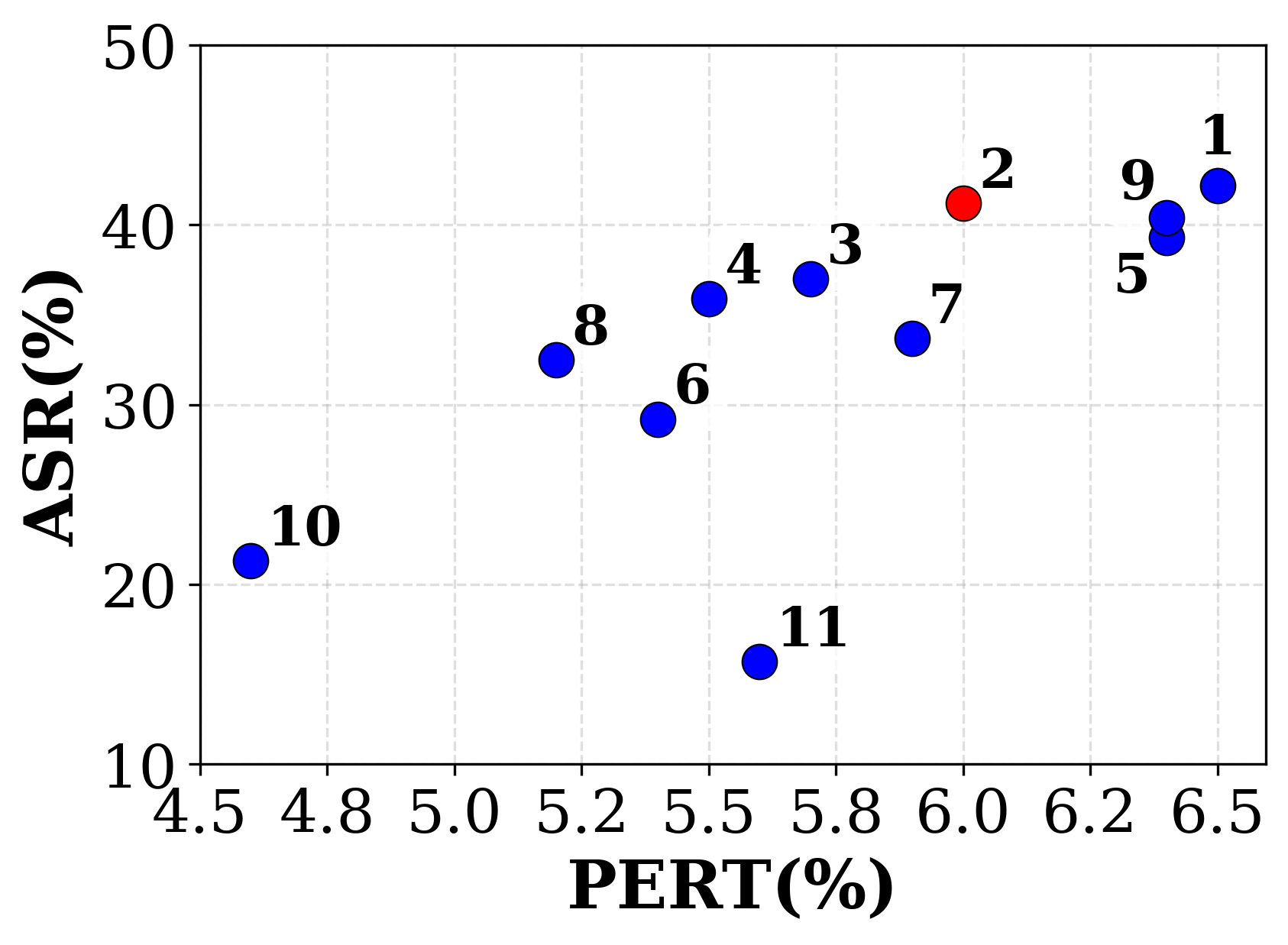}
    \caption{AG (Qwen2.5-FT)}
  \end{subfigure}\hfill
  \begin{subfigure}{0.24\textwidth}
    \centering
    \includegraphics[width=\linewidth]{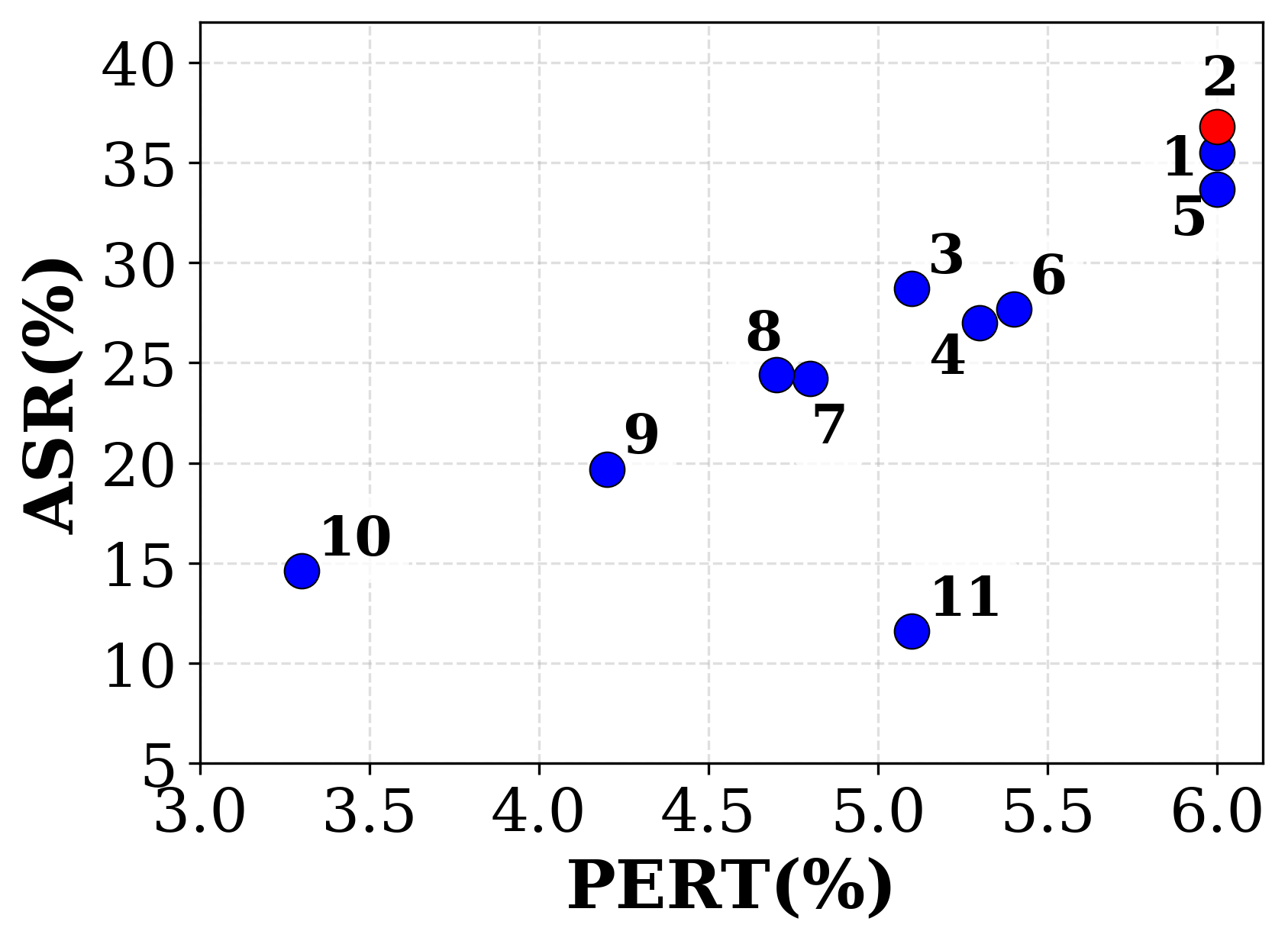}
    \caption{AG (DistilBERT)}
  \end{subfigure}
  \caption{Transfer performance of the 11 Pareto-front candidate chains across Yahoo, Yelp, SST-2, and AG on Qwen2.5-FT and DistilBERT. Higher points indicate higher ASR, and horizontal position indicates Pert. Solution 2 (red) is the TOPSIS-selected chain.}
  \label{fig:topsis}
\end{figure*}

To assess whether the fixed global chain selected on the local dataset under the local model transfers beyond its local search condition, we evaluate all 11 Pareto-front candidate chains across target datasets and target models. Figure~\ref{fig:topsis} collects eight transfer plots on Yahoo, Yelp, SST-2, and AG for Qwen2.5-FT and DistilBERT. Across these target settings, the TOPSIS-selected chain (Solution 2) stays on or near the favorable ASR--Pert frontier.

This pattern follows directly from TOPSIS: after normalizing ASR and Pert, it selects the Pareto-front chain nearest to the ideal point of high ASR and low Pert, rather than an endpoint optimized for a single objective. Equivalently, TOPSIS favors the chain with the smallest joint shortfall on the two objectives, so the selected compromise chain preserves a balanced margin on both metrics and is less sensitive to model and dataset shifts.

\subsection{Comparison with Manually Constructed Chains (RQ5)}
Table~\ref{tab:composition} compares {\OURS} with two manually specified chains on SST-2 and Yahoo using Qwen2.5-FT: \textit{best-3} (the manually constructed chain formed by the top-3 standalone attackers on the local dataset under the local model; HQAAttack $\to$ TextHoaxer $\to$ SSPAttack), \textit{random-3} (a randomly constructed 3-attacker chain; SSPAttack $\to$ LimeAttack $\to$ VIWHard). {\OURS} achieves higher ASR than both manually constructed chains on both datasets. In particular, on Yahoo, {\OURS} surpasses \textit{best-3} by 7.0\%, although its Pert is slightly higher. This is because the additional successfully attacked samples are harder cases, which often require perturbing more words to flip the prediction.

\begin{table}[htbp!]
\centering
\small
\setlength{\tabcolsep}{4pt}
\renewcommand{\arraystretch}{1.05}
\begin{tabular}{ll cc cc}
\toprule
\multirow{2}{*}{\textbf{Model}} & \multirow{2}{*}{\textbf{Method}}
& \multicolumn{2}{c}{\textbf{SST-2}}
& \multicolumn{2}{c}{\textbf{Yahoo}} \\
\cmidrule(lr){3-4} \cmidrule(lr){5-6}
& & \textbf{ASR}$\uparrow$ & \textbf{Pert}$\downarrow$
  & \textbf{ASR}$\uparrow$ & \textbf{Pert}$\downarrow$ \\
\midrule
\multirow{4}{*}{\textbf{Qwen2.5-FT}}
& \textbf{\OURS}          & \textbf{42.0} & 6.8 & \textbf{56.5} & 4.4 \\
& best-3                  & 40.1 & 6.7 & 49.5 & 4.1 \\
& random-3                & 34.5 & \textbf{6.4} & 42.6 & \textbf{3.6} \\
\bottomrule
\end{tabular}
\caption{Comparison with manually constructed baselines on Qwen2.5-FT.}
\label{tab:composition}
\end{table}

\subsection{Ablation Study (RQ6)}
To isolate the contribution of the three components introduced in Sec.~3, we compare {\OURS} with three reduced variants on SST-2 and Yahoo using Qwen2.5-FT: (1) \textit{w/o multi-objective}, which replaces the bi-objective score in Eq.~\eqref{eq:search_objective} with ASR-only search and selects the best-ASR chain (VIWHard $\to$ HQAAttack $\to$ TextHoaxer); (2) \textit{w/o NSGA-II}, which keeps the same bi-objective evaluation and final selector but replaces NSGA-II with random search in the same chain space (VIWHard $\to$ TextHoaxer $\to$ SSPAttack); and (3) \textit{w/o TOPSIS}, which keeps the bi-objective NSGA-II search but replaces the final compromise selector with 10 randomly selected Pareto-front solutions and reports their average performance. As shown in Table~\ref{tab:ablation}, removing any of the three components hurts performance. On Yahoo, the drops are much larger: ASR falls from 56.5 for {\OURS} to 38.8 without TOPSIS and further to 27.3 and 28.1 without multi-objective optimization or NSGA-II-based search. Notably, these degraded variants often achieve lower Pert, because they tend to succeed only on relatively easy samples, and such fragile samples can often be flipped by changing only a few words, even at random positions. 

\begin{table}[htbp!]
\centering
\small
\setlength{\tabcolsep}{4pt}
\renewcommand{\arraystretch}{1.05}
\begin{tabular}{ll cc cc}
\toprule
\multirow{2}{*}{\textbf{Model}} & \multirow{2}{*}{\textbf{Method}}
& \multicolumn{2}{c}{\textbf{SST-2}}
& \multicolumn{2}{c}{\textbf{Yahoo}} \\
\cmidrule(lr){3-4} \cmidrule(lr){5-6}
& & \textbf{ASR}$\uparrow$ & \textbf{Pert}$\downarrow$
  & \textbf{ASR}$\uparrow$ & \textbf{Pert}$\downarrow$ \\
\midrule
\multirow{4}{*}{\textbf{Qwen2.5-FT}}
& \textbf{\OURS}             & \textbf{42.0} & 6.8 & \textbf{56.5} & 4.4 \\
& w/o multi-objective        & 40.8 & 6.8 & 27.3 & 3.4 \\
& w/o NSGA-II    & 41.2 & \textbf{6.7} & 28.1 & \textbf{3.3} \\
& w/o TOPSIS                 & 37.1 & 7.2 & 38.8 & 3.8 \\
\bottomrule
\end{tabular}
\caption{Component ablation results on Qwen2.5-FT.}
\label{tab:ablation}
\end{table}

\begin{figure}[t]
  \centering
  \includegraphics[width=0.72\linewidth]{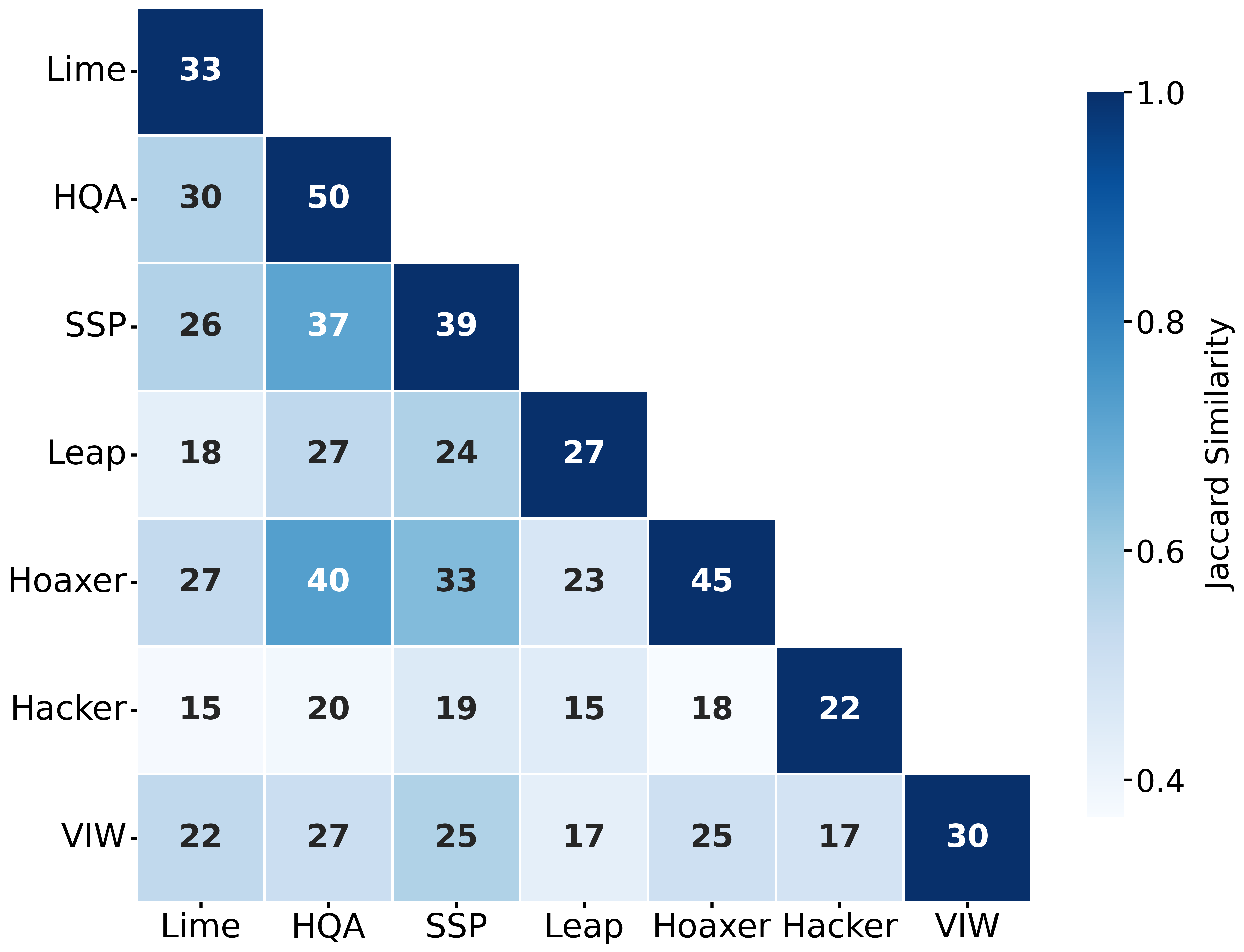}
  \caption{Success-set overlap on local data/model ($N=100$). Diagonal: successful samples; darker color: stronger pairwise overlap.}
  \label{fig:overlap}
\end{figure}

We further analyze attacker complementarity with the overlap matrix in Figure~\ref{fig:overlap}, computed on the local dataset under the local model using $N=100$ samples. The diagonal entries are the numbers of successfully attacked samples for each attacker, while off-diagonal colors indicate pairwise overlap intensity (darker means stronger overlap). Notably, the three attackers in \textit{best-3}, namely HQAAttack, TextHoaxer, and SSPAttack, all show relatively dark overlap cells with one another in the matrix. This suggests that simply combining the strongest standalone attackers may provide limited additional coverage. In contrast, the searched chain can better capture complementarity across different attackers, which helps explain why sequence-level optimization is beneficial.

\subsection{Case Study}
\label{app:case-study}
Table~\ref{tab:case-study} presents four representative DistilBERT examples on SST-2 under the selected chain. A clear pattern emerges: earlier attackers often fail either because they do not flip the prediction or because their candidates violate the unified quality constraints, while later attackers can still find valid adversarial examples for the same input. This shows that different attackers cover different residual cases, and that the gain of the chain comes from cross-attacker complementarity rather than from any single attacker alone.

\begin{table}[t]
\centering
\scriptsize
\setlength{\tabcolsep}{3pt}
\renewcommand{\arraystretch}{1.2}
\begin{tabular}{l c c p{0.60\textwidth}}
\toprule
\textbf{Method} & \textbf{Label} & \textbf{Constraint} & \textbf{Adversarial Text} \\
\midrule
HQAAttack & \cmark & & ... \rb{harshly} `` shakes the clown '' , a much funnier film with a similar theme and an equally great robin williams performance ... \\
VIWHard   & & & ... \rb{roughly} `` shakes the clown '' , a much funnier film with a similar theme and an equally great robin williams performance ... \\
\textbf{\OURS} & & & ... \rb{roughly} `` shakes the clown '' , a much funnier film with a similar theme and an equally great robin williams performance ... \\
\midrule
HQAAttack & & \cmark & \rb{harmoniously} speaking we 're in all of me realm again and strictly speaking schneider is no steve martin \\
VIWHard   & & & \rb{smoothly} speaking we 're in all of me territory again and strictly speaking schneider is no steve martin \\
\textbf{\OURS} & & & \rb{smoothly} speaking we 're in all of me territory again and strictly speaking schneider is no steve martin \\
\midrule
HQAAttack & \cmark & & \rb{smoothly} found in its ability to spoof both black and white stereotypes equally. \\
VIWHard   & \cmark & & \rb{gently} found in its ability to spoof both black and white stereotypes equally. \\
\textbf{\OURS} & & & \rb{warmly} found in its ability to spoof both black and white stereotypes equally. \\
\midrule
HQAAttack & & \cmark & \rb{harshly} are now two signs that m. night shyamalan 's debut feature sucked up all he has to give to the mystic genres of cinema: unbreakable and signs. \\
VIWHard   & & \cmark & \rb{roughly} are now two signs that m. night shyamalan 's debut feature sucked up all he has to give to the mystic genres of cinema: unbreakable and signs. \\
\textbf{\OURS} & & & \rb{coldly} are now two signs that m. night shyamalan 's debut feature sucked up all he has to give to the mystic genres of cinema: unbreakable and signs. \\
\bottomrule
\end{tabular}
\caption{Case study on DistilBERT (SST-2) using the selected chain \textbf{HQAAttack $\to$ VIWHard $\to$ TextHacker}. A checkmark under \textbf{Label} indicates that the prediction is unchanged, while a checkmark under \textbf{Constraint} indicates that the candidate is rejected by the unified quality constraints.}
\label{tab:case-study}
\end{table}

\section{Conclusion}
We revisit hard-label black-box text attacks from the perspective of attacker composition rather than single-attacker design. We show that different attackers follow different search trajectories and exhibit non-trivial complementarity. To address this, {\OURS} formulates attacker composition as an attacker-sequence optimization problem and searches for a fixed reusable global chain. Across datasets, victim models, and large language models, the searched chain consistently outperforms strong standalone baselines and simple manual compositions. Notably, it exposes the vulnerability of Large Language Models (e.g., Qwen2.5) in both zero-shot and robust fine-tuned settings.

\bibliographystyle{splncs04}
\bibliography{custom}
\end{document}